\documentclass[journal]{IEEEtran}
\usepackage{graphicx}
\usepackage{epsfig}
\usepackage{algorithm}
\usepackage[noend]{algorithmic}
\usepackage{amsmath}
\usepackage{url}
\usepackage{color}
\usepackage{subfig}
\usepackage{multicols}
\usepackage{tikz}

\usepgflibrary{shapes}
\usepgflibrary{arrows}

\begin{document}

\title{Path Computation for Provisioning in Multi-Technology Multi-Layer Transport Networks}

%\author{\IEEEauthorblockN{Madanagopal Ramachandran, Krishna M. Sivalingam}\\
%\IEEEauthorblockA{Department of Computer Science and Engineering\\
%Indian Institute of Technology Madras, Chennai, India\\
%and\\
%India UK Advanced Technology Center of Excellence in Next Generation Networks, Systems and Services (IU-ATC)\\
%Email: madan@lantana.tenet.res.in,skrishnam@cse.iitm.ac.in}}

%\author{R.~Madanagopal,
%        and~Krishna~M.~Sivalingam%

%\thanks{R. Madanagopal is with NMSWorks Software Pvt. Ltd., Chennai, India e-mail: (madan@nmsworks.co.in).}%
%\thanks{Krishna M. Sivalingam is with the Department of Computer Science and Engineering, Indian Institute of Technology, Madras, India e-mail: (skrishnam@cse.iitm.ac.in).}}
%}

\author{\textsf{Madanagopal Ramachandran and Krishna M. Sivalingam
    (Fellow, IEEE)}
  \\ \textit{Dept. of CSE, Indian Institute of Technology Madras,
    Chennai, India}
\\\ Emails: \{cs14d009@smail.iitm.ac.in, madanagopal.r@gmail.com, skrishnam@iitm.ac.in\}
%\\ Emails: \{cs14d009@smail.iitm.ac.in, skrishnam@iitm.ac.in\}
}

%\author{R. Madanagopal, Krishna M. Sivalingam}

\maketitle

%\noindent Paper \#110, 13 pages

\begin{abstract}
Service providers employ different transport technologies like PDH,
SDH/SONET, OTN, DWDM, Ethernet, MPLS-TP etc. to support different types
of traffic and service requirements. Dynamic service provisioning requires
the use of on-line algorithms that automatically compute the path to be taken
to satisfy the given service request. A typical transport network element
supports adaptation of multiple technologies and multiple layers of those
technologies to carry the input traffic. Further, transport networks are
deployed such that they follow different topologies like linear, ring,
mesh, protected linear, dual homing etc. in different layers. 
Path computation for service requests considering the above factors is the
focus of this work, where a new mechanism for building an auxiliary
graph which models each layer as a node within each network element and
creates adaptation edges between them and also supports creation of
special edges to represent different types of topologies is proposed.
Logical links that represent multiplexing or adaptation are also created
in the auxiliary graph. Initial weight assignment scheme for non-adaptation
edges that consider both link distance and link capacity is proposed and
three dynamic weight assignment functions that consider the
current utilization of the links are proposed. Path computation algorithms
considering adaptation and topologies are proposed over the auxiliary
graph structure. The performance of the algorithms is evaluated and it is
found that the weighted number of requests accepted is higher and the
weighted capacity provisioned is lesser for one of the dynamic weight function
and certain combination of values proposed as part of the weight assignment.
\end{abstract}

\begin{IEEEkeywords}
Multi-Technology Multi-Layer transport networks, Transport SDN, Adaptation across layers, Dedicated protection, Path computation, Provisioning, Shared Risk Link Group (SRLG)
\end{IEEEkeywords}

\section{Introduction}

In the current environment of high speed communications, network operators
and service providers are constantly flooded with bandwidth requirements.
The demand for adding capacity has led to the development of different transport technologies
such as PDH, SDH/SONET, WDM/DWDM, OTN, Ethernet, MPLS-TP etc. which
support different types of traffic requirements. Service providers require
faster provisioning of services to satisfy more customer requests in less time.
Typically, services are provisioned manually by the service providers by
logging into management systems and doing provisioning related activities
using them which is slow as highlighted in \cite{TSDN:janz}.

Automating the service provisioning process will be a key enabler for
satisfying more customer service requests. Automation would result in lesser
time-to-provision customer service requests and it would also help in
efficient usage of network resources. This is a practical problem since
in a large multi-technology transport network comprising tens of thousands
of network elements, there is no single view of the entire network for the
service providers. With the advent of Software Defined Networking (SDN)
and centralized control of the network, automation of service
provisioning is possible.

In Transport SDN, a centralized SDN controller
with a Path Computation Element (PCE) module can automate the path finding
and activation of circuit or services in the transport network. The
SDN controller would be able to discover the network topology
by communicating with the network elements or Network Management Systems (NMS)
using standard or proprietary protocols. It can use the PCE to find paths
and again use the same or different set of protocols to perform configuration or activation
of circuits or services in the network. Path computation can be provided as
a service or API as part of the Open Networking Foundation (ONF) Transport
API (TAPI) which can be used to provide bandwidth-on-demand,
virtual transport network services and also perform multilayer
optimization as mentioned in \cite{TSDN:janz}.

Path computation plays a major role in automating the service provisioning
process in large transport networks which uses multiple technologies.
Further, each transport technology will have multiple layers (ex. in SDH - VC12,
VC3 service and VC4 bearer, in OTN - ODU-j bearer and OTU-k/OCH link/channel)
in which connection support will be possible and there would be adaptation
from one layer in some technology to another layer in the same or different
technology. The transport networks are typically organized as logical
topologies which ensures easy maintenance of the network,
and better protection and resilience wherever required.

The algorithms to find the best path in a transport network taking into account
the multiple technologies and multiple layers within each technology is the
focus of this work. This requires consideration of constraints in creation of
connections in the layered transport network and adaptation between the layers.
Since the transport networks are organized as logical topologies, path
computation should also consider them for efficient routing. An auxiliary graph structure
which models each technology and layer combination within every network
element as nodes and adaptation between them as links is proposed. 
Similar to the approach followed in \cite{kuipers2009path},
adaptation edges between layers and technologies are created in the auxiliary graph structure. But this
work assumes that the chances of same node or edge getting visited multiple times due to adaptation
constraints is not present, since in a service provider transport network the chances of encountering
such a scenario is very rare due to proper organization of the transport network across technologies
and layers. Due to the way traffic cards corresponding to technology adaptation are available
in a real transport network element which support only way of adaptation, the chances
of a node getting visited multiple times is not practical.
The reason for relaxing that condition is that, it leads to path computation
taking exponential time as mentioned in \cite{kuipers2009path}.
The mechanism for creating logical links and building logical topologies as special edges
in the auxiliary graph is also proposed. Methods for weight assignment for the
physical links and special edges are proposed followed by algorithms for computing
unprotected and link-disjoint path pair (for end-to-end protection). The algorithms
consider Shared Risk Link Groups (SRLGs) present in the transport network also.
This work is aimed at usage for satisfying service requests (typically long-lived)
in a service provider transport network which contains nodes in the order of thousands
and not intended for routed networks with millions of nodes.

%The remainder of this paper is organized as follows. Section 2 gives details
%regarding related work. Section 3 describes the transport network organization.
%Section 4 describes the mechanism for building the auxiliary graph structure.
%Section 5 describes the path computation algorithms for transport networks based on the
%proposed graph structure. Section 6 compares the performance of the path
%computation algorithms for the weight assignment schemes proposed.
%Finally, Section 7 concludes this paper.

\section{Background and Related Work}

\subsection{Transport Network}

A transport network of a typical service provider is deployed such that it uses different
transport technologies for different bandwidth and service requirements. It comprises of a
national long distance (NLD) network that transports traffic from one city to another. It is typically
formed as a mesh network with Automatically Switched Optical Network (ASON) functionality
implemented. To cater to the increasing bandwidth demands, optical fibers supporting WDM
is deployed in the NLD network. One choice for the NLD network is a OTN mesh network
supporting ASON at ODU-k layer which multiplexes multiple OTN signals over WDM. Each city
in the country has one or more OTN nodes which aggregates traffic from the city to transport
over the NLD network to other cities in the country. Other choices for the NLD network are OTN
with OCH ASON (Wavelength Switched Optical Network - WSON) or OTN with Flexgrid which supports
ASON at OCH spectrum layer (Spectrum Switched Optical Network - SSON).
%The typical connection types in a transport network with the client and server layers for each of those connection
%types is listed in Table \ref{conn-types}.

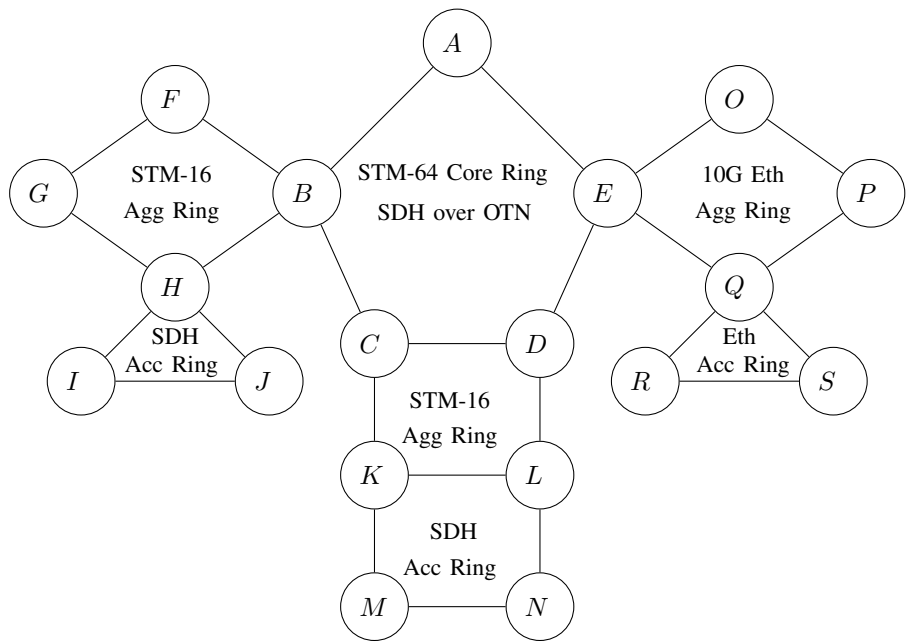
\begin{figure*}
\centering
\resizebox{!}{!}{\begin{tikzpicture}

%\tikzset{
\tikzstyle{networknode} = [
%  networknode/.style = {
    draw,
    shape=circle,
%    radius=2cm,
    minimum width=0.9cm,
    text width=0.4cm
%  }
    ]
%}

\node[networknode] (A) at (0,0) {$A$};
\node[networknode] (B) at (-2,-2) {$B$};
\node[networknode] (C) at (-1.1,-4) {$C$};
\node[networknode] (D) at (1.1,-4) {$D$};
\node[networknode] (E) at (2,-2) {$E$};

\node[networknode] (F) at (-3.75,-0.75) {$F$};
\node[networknode] (G) at (-5.5,-2) {$G$};
\node[networknode] (H) at (-3.75,-3.25) {$H$};

\node[networknode] (I) at (-5,-4.5) {$I$};
\node[networknode] (J) at (-2.5,-4.5) {$J$};

\node[networknode] (O) at (3.75,-0.75) {$O$};
\node[networknode] (P) at (5.5,-2) {$P$};
\node[networknode] (Q) at (3.75,-3.25) {$Q$};

\node[networknode] (S) at (5,-4.5) {$S$};
\node[networknode] (R) at (2.5,-4.5) {$R$};

\node[networknode] (K) at (-1.1,-5.75) {$K$};
\node[networknode] (L) at (1.1,-5.75) {$L$};

\node[networknode] (M) at (-1.1,-7.5) {$M$};
\node[networknode] (N) at (1.1,-7.5) {$N$};

\draw (A) -- (B);
\draw (B) -- (C);
\draw (C) -- (D);
\draw (D) -- (E);
\draw (E) -- (A);

\draw (B) -- (F);
\draw (F) -- (G);
\draw (G) -- (H);
\draw (H) -- (B);

\draw (H) -- (I);
\draw (I) -- (J);
\draw (J) -- (H);

\draw (E) -- (O);
\draw (O) -- (P);
\draw (P) -- (Q);
\draw (Q) -- (E);

\draw (Q) -- (R);
\draw (R) -- (S);
\draw (S) -- (Q);

\draw (C) -- (K);
\draw (K) -- (L);
\draw (L) -- (D);

\draw (K) -- (M);
\draw (M) -- (N);
\draw (N) -- (L);

\node at (-0.06,-1.75) {\small STM-64 Core Ring};
\node at (-0.04,-2.25) {\small SDH over OTN};

\node at (-3.8,-1.75) {\small STM-16};
\node at (-3.8,-2.25) {\small Agg Ring};

\node at (3.8,-1.75) {\small 10G Eth};
\node at (3.8,-2.25) {\small Agg Ring};

\node at (-3.75,-3.9) {\small SDH};
\node at (-3.8,-4.3) {\small Acc Ring};

\node at (3.75,-3.9) {\small Eth};
\node at (3.8,-4.3) {\small Acc Ring};

\node at (-0.1,-4.75) {\small STM-16};
\node at (-0.1,-5.25) {\small Agg Ring};

\node at (-0.05,-6.5) {\small SDH};
\node at (-0.1,-7) {\small Acc Ring};

\end{tikzpicture}}
\caption{Metro Network}
\label{MetroNetwork1}
\end{figure*}

The city or metro network is again a WDM network which transports SDH and Ethernet traffic within the
city. The traffic to other cities are handed over to the NLD aggregation node which is part of the NLD
network. SDH/SONET is used to carry voice and other fixed rate traffic and Ethernet is used to
carry variable rate traffic like IP, IP-MPLS and MPLS-TP etc. The SDH metro network is typically
organized as multiple core rings at STM-64 rate whose nodes are connected by WDM over
optical fibers. Each core ring has multiple aggregate rings typically at STM-16 rate. Each aggregate
ring has multiple access rings typically at STM-4/STM-1 rates. The Ethernet metro network is
also typically organized as multiple core rings at 10 Gbps rate. Each core ring has
multiple aggregate or access rings typically at 1 Gbps rate. In both SDH/SONET and Ethernet metro
networks, the aggregate and access rings could be subtended at one node or at two nodes (dual
homing rings).

A depiction of the metro network is shown in Fig.~\ref{MetroNetwork1}. It contains a core WDM network
with nodes \emph{a}, \emph{b}, \emph{c}, \emph{d} and \emph{e} which are also involved in a SDH core ring
The figure shows a SDH aggregate ring at STM-16 rate involving the nodes \emph{b},
\emph{f}, \emph{g} and \emph{h} with \emph{b} being the aggregate node and a Ethernet aggregate ring
at 10 Gbps rate involving the nodes \emph{e}, \emph{o}, \emph{p} and \emph{q} with \emph{e} being
the aggregate node. The SDH aggregate ring has a access ring at STM-4/STM-1 rates involving the
nodes \emph{h}, \emph{i}, and \emph{j} with \emph{h} being the aggregate node.
The Ethernet aggregate ring has a access ring at 1 Gbps rate, that supports MPLS-TP involving the nodes \emph{q}, \emph{r}
and \emph{s} with \emph{q} being the aggregate node. In SDH case, a dual homing ring involving
the nodes \emph{c}, \emph{d}, \emph{k} and \emph{l} with \emph{c} and \emph{d} being the aggregate nodes
is also shown. This dual homing ring has another dual homing ring involving the nodes \emph{k}, \emph{l},
\emph{m} and \emph{n} with \emph{k} and \emph{l} being the aggregate nodes.

Service provisioning in transport networks is an important process which is performed
on a routine basis. With centralized control of the network using Transport SDN, service providers
would be able to get a complete view of their network. This would enable path
computation for service provisioning using specific applications which implement PCE. 
Path computation for automatic service provisioning has been extensively studied in certain types of
transport networks. In some prior works, the path computation problem has been
considered in multi-layer networks.

\subsection{Path Computation in SONET/SDH Networks}

When it comes to provisioning in transport networks particularly SONET/SDH networks, links cannot
be simply considered to have certain integer units of capacity as the
multiplexing structure defined by those technologies has to be followed while
provisioning the requested services. The multiplexing structure of SONET is
similar to SDH but with different terminologies like Virtual Tributaries (VT)
in SONET instead of Virtual Containers (VC) in SDH.

Because of this multiplexing structure of SDH, each link cannot be assumed
to have simply some integer units of capacity and the free capacity cannot be
obtained simply by subtracting the allotted capacity from the maximum capacity
and allocations cannot be made until the free capacity becomes zero. In SDH,
higher order containers like VC-4 have to be used to create trails (logical
connections) between the source and destination nodes before provisioning
any bandwidth between two points.

Path Computation algorithms taking into account the above constraints have been
proposed in \cite{PCASDH:madan}. The network is treated as a graph containing
physical links and logical trails and weights are assigned to them before
computing a path with the least cost. Weights are assigned such that the trails
are given higher preference to physical links.
%so that existing trails are used wherever possible.

In \cite{PCADPSPIM:madan}, path computation algorithms for dynamic service
provisioning with protection and inverse multiplexing in SONET/SDH networks
are proposed. In dedicated protection, standard protection mechanisms defined
in SONET/SDH like Multiplex Section Protection (MSP), Multiplex Section - Shared
Protection Ring (MS-SPRing) and Subnetwork Connection Protection (SNCP) are
taken into account. In shared protection, a minimum information scenario is
proposed as a trade-off between maintaining less sharing information and
maximizing sharing.

The algorithms to find the best path in a transport network taking into account
the different topologies deployed is proposed in \cite{TPCA:madan} where
two types of graphs are considered, a super graph containing the topologies as
nodes and common nodes as edges and the other, the base graph containing network
elements as nodes and physical links as edges. Path computation is done in the
super graph first where the path is found in terms of topologies and then the
basic path is found in the graphs representing the topologies. The algorithms
proposed consider no protection and dedicated protection. Even though
\cite{TPCA:madan} considers topology constraints, it is applicable for a specific
technology (like SDH alone) and not in a multi-technology multi-layer setting.

\subsection{Path Computation in Multi-Layer Networks}

A comprehensive survey of the challenges related to network management in multi-layer
multi-vendor networks is provided in \cite{martinez2014network} where the earlier
work on multi-layer path computation algorithms is mentioned. It classifies the
related work based on constraint type (prunable and non-prunable) for path
computation and provides the list of references to the proposed solutions to the problem.

In \cite{kuipers2009path}, two models of graph for multi-layer networks, one which
models each layer supported in a network element as a node and adaptations
between the layers as edges, and the other which models each technology stack
as a node and edges between them to represent direct communication or adaptation.
The problem of finding paths in multi-layer networks is shown to be NP-complete and
path selection algorithms for both the models are proposed. The algorithms consider
the scenario where a node and hence an edge could be visited multiple times.

\begin{table*}
\begin{center}
\captionof{table}{Summary of contributions}
\label{contributions}
\begin{tabular}{|l|l|l|}
\hline
Design Decision & Rationale/Benefits & Challenges\\
\hline
Cross-Layer auxiliary graph structure & 1. Gives better control of optimizing & 1. Scalability\\
& the network resources & 2. Loss of simplicity in \\
& 2. Avoids hierarchical layer-wise graph & layered graph\\
& which could be sub-optimal & \\
\hline
Special edges for topologies in graph & 1. Working and Protection paths & 1. Increase in the number \\
& follow the same set of topologies & of edges in graph \\
& 2. Helps in easy maintenance and troubleshooting & \\
& 3. Avoids misaligned working and protection & \\
& paths with respect to topologies & \\
\hline
Initial weight assignment for edges & 1. Balanced utilization & 1. Evaluation required to find \\
considering both distance and capacity & 2. Proportional resource assignment & the right balance \\
\hline
Dynamic weight assignment for & 1. Preference for lesser utilization & 1. Finding the right thresholds\\
edges based on utilization & 2. Avoids over-utilization & for increasing dynamic weight \\
\hline
Heuristic Path Computation & 1. Supports protected path computation & 1. Computation time \\
Algorithms & 2. Finds, creates and uses logical links & \\
& 3. Supports compatible layer adaptations & \\
\hline
\end{tabular}
\end{center}
\end{table*}

A multi-layer network model is proposed in \cite{dijkstra2008multi} which is based on the ITU-T
G.805 \cite{G.805:itut} functional elements. Path finding in this scenario is shown to be
a path-constrained problem instead of link-constrained problem and algorithm
using breadth-first search is proposed in \cite{dijkstra2009path} to compute the
best path using that model.

In \cite{iqbal2015technology}, modeling each domain in a multi-domain multi-layer optical network as
technology matrix and each inter-domain link as technology vector with costs associated
with them based on different parameters, is proposed. The path computation problem using this
modeling is shown to be NP-hard and path finding algorithms based on the proposed
modeling to compute exact paths in a multi-domain network is proposed.

%The problem of path finding in multi-domain networks with technology
%incompatibilities is proposed in \cite{iqbaltechnology}. The problem is shown
%to be HP-hard and path finding algorithm which uses k-shortest paths
%approach by maintaining a list of feasible subpaths at each intermediate
%domain is proposed.

Path computation for multi-layer networks with and without consideration of
bandwidth constraints and under QoS constraints is studied in \cite{PCMLAlgoComp:Lamali}.
Polynomial algorithms for path computation without bandwidth constraint is
proposed by building a Push Down Automata (PDA) for modeling the multi-layer
network, converting it into a Context Free Grammar (CFG) and computing the
shortest word generated by the CFG. The same problem with bandwidth constraint is
shown to be NP-Complete and heuristic algorithm is proposed.

All the approaches mentioned in this subsection consider computation of only one
path between a given pair of nodes and as such support for link-disjoint path pair computation
which is required for protection is not mentioned. Also, they do not consider initial and
dynamic weight assignment which is required for balancing the utilization in the network.
These two aspects are considered in this work.

\section{Contributions}

The summary of contributions of this work where the design decisions made along with the
reasons, benefits and challenges are listed in the Table \ref{contributions}.
Path computation in transport networks has to consider the different technologies
and the layers supported in each of those technologies. Also, for path computation
to be practically used in service provider's transport network, topologies like
ring, dual homing, linear etc. have to be considered so that the provisioned path
is easy to troubleshoot and maintain. Troubleshooting refers to the circuit
level analysis due to link failures in the network and not just link failures.
Ring is a topology where the links form a cycle such that it originates
and terminates in the same node which is the aggregation node through which
all traffic from/to the ring enters/exits. Dual Homing is a topology
which is linear and has two aggregate nodes at the ends
of the linear section. In this case, for all traffic from/to the dual homing
section, working path has to enter/exit from one aggregation node and the protection
path has to enter/exit from the other aggregation node. Mesh is a topology
where the edges are arbitrarily connected between the nodes.
Within the topology, the shortest path between the involved nodes in the
topology would be computed as the working path and the longer
path would be computed as the protection path.
Path computation also has to consider the different factors
like link distance and utilization for efficient usage of the network so that
more requests can be provisioned in the future. These aspects are not considered
in the earlier works on path computation in multi-technology multi-layer
transport networks to the best of our knowledge. In this work, these factors
are considered by proposing an auxiliary graph with support for physical, logical,
adaptation edges and special edges to consider topologies. 
While computing path in the multi-layer auxillary graph, any
adaptation that is done from one layer to another layer has to be done
in the reverse manner in some other node later in the path, so that
the traffic part of the service request would really flow in the data plane.
Initial and dynamic weight assignment schemes for links are proposed to balance the utilization
and accommodate more future requests. Path computation in transport networks have to be done such that the output
route is typically able to withstand failures like cable/fiber cut (link failure).
This is achieved by computing two paths - working and protection such that they
are edge-disjoint. The algorithms for unprotected and protected
path computation in the auxiliary graph is proposed and their performance is
evaluated for the proposed weight assignment schemes.
Links between transport network elements are typically carried over cables which
contain more number of such links. When a cable gets cut, all the links get
affected and if both the working and protection path for a service happen to
pass through the cable, then the service will be impacted and failed.
To avoid this, logical grouping of links (physical or logical) are modelled as
Shared Risk Link Group (SRLG) where the involved links have a common risk of failure.
While finding protection path part of edge-disjoint path pair, the links
involved in the SRLGs part of the working path are excluded so that
working and protection combination is more resilient to cable cuts.

\section{Auxiliary Graph Construction}

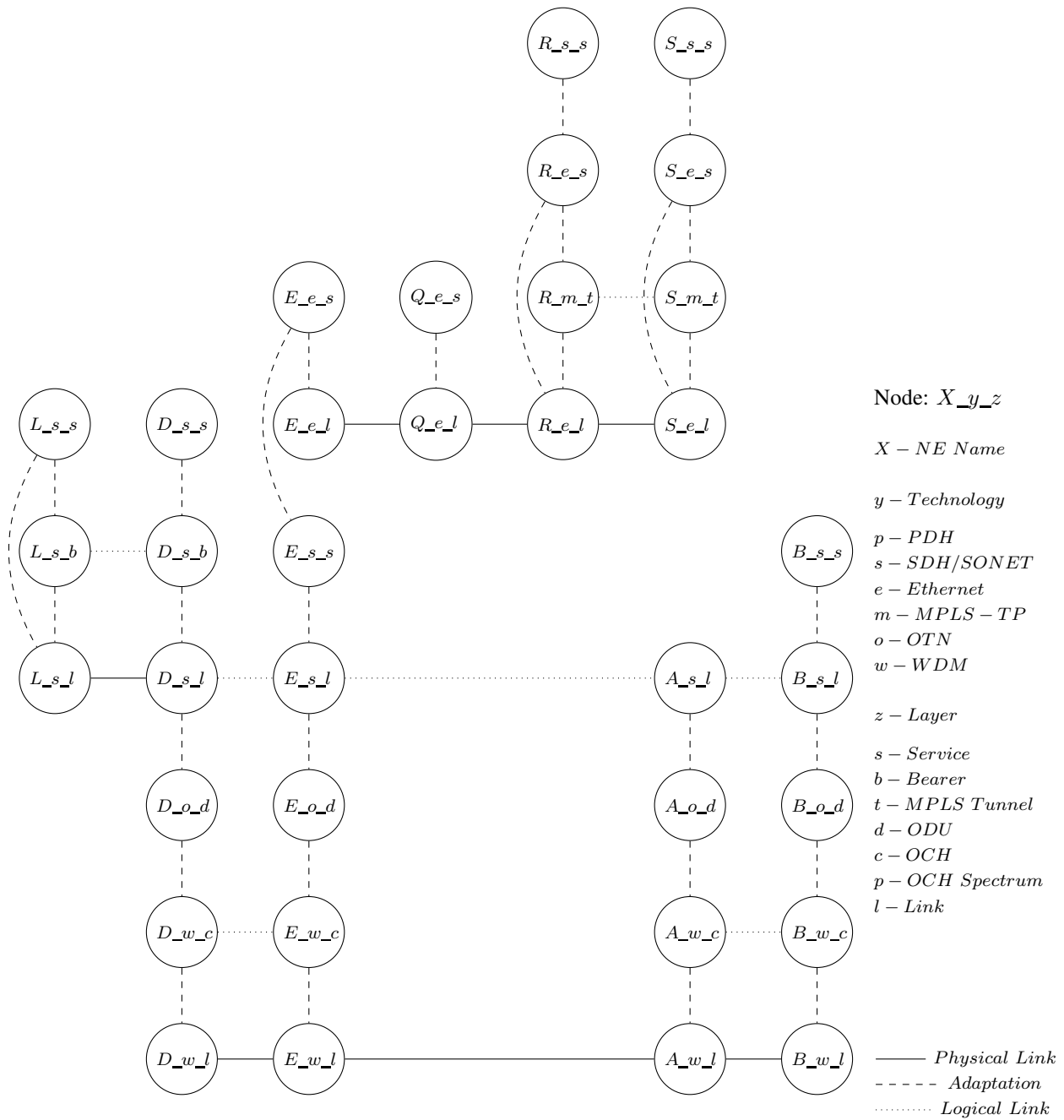
\begin{figure*}
\centering
\resizebox{!}{!}{\begin{tikzpicture}[node distance = 2cm,auto]

%\tikzset{
\tikzstyle{networknode} = [
%  networknode/.style = {
    draw,
    shape=circle,
%    radius=2cm,
    minimum width=0.9cm,
    text width=0.8cm
%  }
    ]
%}

\tikzstyle{mytext}=[draw=none, 
  minimum width=3.2cm,text width=3cm]

%\DrawCircle{0}{0}{thick}{14pt}{black}{M}
%\DrawCircle{20}{0}{thick}{14pt}{black}{G}
%\DrawNormalLine{0}{10}{0}{12}{thick}
%\DrawDashedLine{1}{10}{1}{12}{thick}
%\DrawDottedLine{2}{10}{2}{12}{thick}

\node[networknode] (Rss) at (0, 0) {\footnotesize $R\_s\_s$};
\node[networknode, below of = Rss] (Res) {\footnotesize $R\_e\_s$};
\node[networknode, below of = Res] (Rmt) {\footnotesize $R\_m\_t$};
\node[networknode, below of = Rmt] (Rel) {\footnotesize $R\_e\_l$};
\node[networknode, right of = Rss] (Sss) {\footnotesize $S\_s\_s$};
\node[networknode, below of = Sss] (Ses) {\footnotesize $S\_e\_s$};
\node[networknode, below of = Ses] (Smt) {\footnotesize $S\_m\_t$};
\node[networknode, below of = Smt] (Sel) {\footnotesize $S\_e\_l$};
\node[networknode, left of = Rmt] (Qes) {\footnotesize $Q\_e\_s$};
\node[networknode, below of = Qes] (Qel) {\footnotesize $Q\_e\_l$};
\node[networknode, left of = Qes] (Ees) {\footnotesize $E\_e\_s$};
\node[networknode, below of = Ees] (Eel) {\footnotesize $E\_e\_l$};
\node[networknode, below of = Eel] (Ess) {\footnotesize $E\_s\_s$};
\node[networknode, below of = Ess] (Esl) {\footnotesize $E\_s\_l$};
\node[networknode, below of = Esl] (Eod) {\footnotesize $E\_o\_d$};
\node[networknode, below of = Eod] (Ewc) {\footnotesize $E\_w\_c$};
\node[networknode, below of = Ewc] (Ewl) {\footnotesize $E\_w\_l$};
\node[networknode, left of = Eel] (Dss) {\footnotesize $D\_s\_s$};
\node[networknode, below of = Dss] (Dsb) {\footnotesize $D\_s\_b$};
\node[networknode, below of = Dsb] (Dsl) {\footnotesize $D\_s\_l$};
\node[networknode, below of = Dsl] (Dod) {\footnotesize $D\_o\_d$};
\node[networknode, below of = Dod] (Dwc) {\footnotesize $D\_w\_c$};
\node[networknode, below of = Dwc] (Dwl) {\footnotesize $D\_w\_l$};
\node[networknode, left of = Dss] (Lss) {\footnotesize $L\_s\_s$};
\node[networknode, below of = Lss] (Lsb) {\footnotesize $L\_s\_b$};
\node[networknode, below of = Lsb] (Lsl) {\footnotesize $L\_s\_l$};
\node[networknode, below of = Sel, right of = Sel] (Bss) {\footnotesize $B\_s\_s$};
\node[networknode, below of = Bss] (Bsl) {\footnotesize $B\_s\_l$};
\node[networknode, below of = Bsl] (Bod) {\footnotesize $B\_o\_d$};
\node[networknode, below of = Bod] (Bwc) {\footnotesize $B\_w\_c$};
\node[networknode, below of = Bwc] (Bwl) {\footnotesize $B\_w\_l$};
\node[networknode, left of = Bsl] (Asl) {\footnotesize $A\_s\_l$};
\node[networknode, below of = Asl] (Aod) {\footnotesize $A\_o\_d$};
\node[networknode, below of = Aod] (Awc) {\footnotesize $A\_w\_c$};
\node[networknode, below of = Awc] (Awl) {\footnotesize $A\_w\_l$};

\draw[dashed] (Rss) -- (Res);
\draw[dashed] (Res) -- (Rmt);
\draw[dashed] (Rmt) -- (Rel);
\draw[dashed] (Sss) -- (Ses);
\draw[dashed] (Ses) -- (Smt);
\draw[dashed] (Smt) -- (Sel);
\draw[dashed] (Qes) -- (Qel);
\draw[dashed] (Ees) -- (Eel);
\draw[dashed] (Ess) -- (Esl);
\draw[dashed] (Esl) -- (Eod);
\draw[dashed] (Eod) -- (Ewc);
\draw[dashed] (Ewc) -- (Ewl);
\draw[dashed] (Dss) -- (Dsb);
\draw[dashed] (Dsb) -- (Dsl);
\draw[dashed] (Dsl) -- (Dod);
\draw[dashed] (Dod) -- (Dwc);
\draw[dashed] (Dwc) -- (Dwl);
\draw[dashed] (Lss) -- (Lsb);
\draw[dashed] (Lsb) -- (Lsl);
\draw[dashed] (Bss) -- (Bsl);
\draw[dashed] (Bsl) -- (Bod);
\draw[dashed] (Bod) -- (Bwc);
\draw[dashed] (Bwc) -- (Bwl);
\draw[dashed] (Asl) -- (Aod);
\draw[dashed] (Aod) -- (Awc);
\draw[dashed] (Awc) -- (Awl);

\draw[dotted] (Rmt) -- (Smt);
\draw[dotted] (Lsb) -- (Dsb);
\draw[dotted] (Dsl) -- (Esl);
\draw[dotted] (Asl) -- (Bsl);
\draw[dotted] (Esl) -- (Asl);
\draw[dotted] (Dwc) -- (Ewc);
\draw[dotted] (Awc) -- (Bwc);

\draw (Rel) -- (Sel);
\draw (Eel) -- (Qel);
\draw (Qel) -- (Rel);
\draw (Lsl) -- (Dsl);
\draw (Dwl) -- (Ewl);
\draw (Awl) -- (Bwl);
\draw (Ewl) -- (Awl);

\draw[bend right, dashed] (Res) to (Rel);
\draw[bend right, dashed] (Ses) to (Sel);
\draw[bend right, dashed] (Ees) to (Ess);
\draw[bend right, dashed] (Lss) to (Lsl);

%\draw[->] (M)  -- (G);

\node[mytext, right of = Bss, above of = Bss,  node distance = 2.4cm] (NodeDesc) {Node: $X\_y\_z$};
\node[mytext, below of = NodeDesc, node distance = 0.8cm] (X) {\footnotesize $X - NE \hspace{0.1cm} Name$};
\node[mytext, below of = X, node distance = 0.8cm] (y) {\footnotesize $y - Technology$};
\node[mytext, below of = y, node distance = 0.6cm] (pdh) {\footnotesize $p - PDH$};
\node[mytext, below of = pdh, node distance = 0.4cm] (sdh) {\footnotesize $s - SDH/SONET$};
\node[mytext, below of = sdh, node distance = 0.4cm] (eth) {\footnotesize $e - Ethernet$};
\node[mytext, below of = eth, node distance = 0.4cm] (mplstp) {\footnotesize $m - MPLS-TP$};
\node[mytext, below of = mplstp, node distance = 0.4cm] (otn) {\footnotesize $o - OTN$};
\node[mytext, below of = otn, node distance = 0.4cm] (wdm) {\footnotesize $w - WDM$};
\node[mytext, below of = wdm, node distance = 0.8cm] (z) {\footnotesize $z - Layer$};
\node[mytext, below of = z, node distance = 0.6cm] (service) {\footnotesize $s - Service$};
\node[mytext, below of = service, node distance = 0.4cm] (bearer) {\footnotesize $b - Bearer$};
\node[mytext, below of = bearer, node distance = 0.4cm] (tunnel) {\footnotesize $t - MPLS \hspace{0.1cm} Tunnel$};
\node[mytext, below of = tunnel, node distance = 0.4cm] (odu) {\footnotesize $d - ODU$};
\node[mytext, below of = odu, node distance = 0.4cm] (och) {\footnotesize $c - OCH$};
\node[mytext, below of = och, node distance = 0.4cm] (spectrum) {\footnotesize $p - OCH \hspace{0.1cm} Spectrum$};
\node[mytext, below of = spectrum, node distance = 0.4cm] (link) {\footnotesize $l - Link$};

%\node[below of = link, node distance = 0.8cm] (blank1) {};
\node[right of = Bwl, node distance = 0.8cm] (blank1) {};
\node[right of = blank1] (physicallink) {\footnotesize $Physical \hspace{0.1cm} Link$};
\draw (blank1) -- (physicallink);
\node[below of = blank1, node distance = 0.4cm] (blank2) {};
\node[right of = blank2] (adaptation) {\footnotesize $Adaptation$};
\draw[dashed] (blank2) -- (adaptation);
\node[below of = blank2, node distance = 0.4cm] (blank3) {};
\node[right of = blank3] (logicallink) {\footnotesize $Logical \hspace{0.1cm} Link$};
\draw[dotted] (blank3) -- (logicallink);
 
\end{tikzpicture}}
\caption{Multi-Layer Graph corresponding to a part of the Metro Network}
\label{MultiLayerGraph}
\end{figure*}

For computing a path for service request in a transport network, a graph that captures the
important characteristics of the network is necessary. The graph can be built in many ways
like having a single graph for the entire network or a hierarchical graph with multiple levels
such that each layer or technology is abstracted in the level above.

%This work proposes a new mechanism for building the graph of the transport network such that the following requirements are met:

%\begin{enumerate}
%\item{It should contain all the nodes and physical links.}
%\item{It should have support for finding and creating logical links which are available when a
%connection layer or technology is carried over another layer or technology.}
%\item{It should have support for logical topologies like linear, ring, mesh, protected linear,
%dual homing etc. which are typically formed out of physical or logical links to provide
%some type of protection or to avoid single point of failure.}
%\item{It should support identification of inter technology links.}
%\item{It should support weight assignment for edges like physical and logical links that
%reflects different parameters (ex. distance, bandwidth utilization etc.) for finding a suitable path
%for service requests.}
%\end{enumerate}

In this work, a single graph of the entire network is built instead of hierarchical graph with multiple levels
since a single graph will give better control of optimizing the network resources. For example, for a service
request between two nearby nodes, the path computation in the hierarchical graph would try to find the path
in the given layer (ex. SDH VC4) graph. If a direct path is available in that layer graph, it would be used.
Else, a path in the same layer graph which could be long and consume more bandwidth would be computed since
it would try to use the available edges in that graph to the extent possible. Only when there is no path
available in that layer graph, the path computation would try to use the server layer (ex. OTN ODU).
Similarly, to satisfy a service request between two farther locations, there would be two different types of
paths, one which could involve a large number of same technology edges but with lesser bandwidth availability
and the other which involves different technologies but with large bandwidth availability. In this case, it
would be better to use the path involving different technologies and larger bandwidth since using the
large number of same technology edges might not be always optimal for accepting more future requests. These problems could be avoided in the
single graph approach by suitably choosing weights for the edges that reflects distance, bandwidth consumed etc.
%which would result in a better path.

The auxiliary graph is constructed as follows:

\begin{enumerate}
\item{All the network elements that support different transport technologies are added
to the graph such that for each network element, nodes that represent different technology and layer combination that
is supported by that network element are added to the graph. Modeling one node for each technology and layer
combination is done assuming connection or switching or adaptation flexibility at each layer.}
\item{For each network element, edges that represent adaptation are added between nodes corresponding to the
supported adaptation type between two technology and layer combination.
Typical adaptations that are possible in a transport network are SDH/SONET/Ethernet logical link
over OTN (ODU-j) or WDM (OTU-k), OTN (ODU-j) over WDM (Lambda Channel or FlexGrid Spectrum), Ethernet Service
over MPLS Tunnel, MPLS Tunnel over Ethernet link etc.}
%These adaptation links correspond to
%one of the client to server adaptation types given in Table \ref{conn-types}.}
\item{All the physical links are added as edges between the nodes.}
%For each link, the link type (ex. PDH-PDH,
%SDH-SDH, SDH-SDHTRANSBW, TRANSCOL-ONELAMBDA, ETHERNET-ETHERNET, ETHERNET-ETHERNETTRANSBW, NLAMBDA-WDM,
%WDM-WDM etc.) is maintained as an attribute.}
%\item{For each physical link, the interTechnology flag is set as true when that physical link is used to
%carry traffic from one technology to another technology (ex. SDH-SDHTRANSBW, SDH-ONELAMBDA, ETHERNET-ETHERNETTRANSBW,
%ETHERNET-ONELAMBDA etc.)}
\item{For each physical link that spans across technologies, a trace of the physical links starting from
the source port of the link is made to check whether it terminates in a destination port of another
inter-technology link. If the tracing is successful, then a logical link between the source and destination
node corresponding to the layer in the network element is added to the graph. The physical and logical links in a network element connect the node corresponding
to the layer in which the link is operating.}
\item{All the logical topologies in the network are taken and considered for addition into the graph with the
following steps:}
\begin{enumerate}
\item{If the topology type is ring and an aggregate node (node through which traffic passes to the ring) exists for that topology (not a core ring), then
special edges are added from each node in that topology to the aggregate node with self-protected flag set to true.
Self-protected flag is used to indicate that the edge corresponding to the
topology is completely protected and hence working and protection paths have to
be found within the topology.}
\item{If the topology type is ring and no aggregate node exists for that topology (core ring), then
full mesh of edges between every pair of nodes in that topology are added with self-protected flag set to true.}
\item{If the topology type is dual homing, a special hub node is added to the graph and special edges are added from
each node in that topology to the special hub node. From the special hub node, two edges to the two aggregate nodes
for that dual homing topology are added to the graph.}
\item{If the topology type is mesh, then the physical or logical links between the nodes in that mesh topology
are already added in the graph.}
%are tagged with the attribute isInvolvedInMesh set as true.}
\end{enumerate}
\end{enumerate}

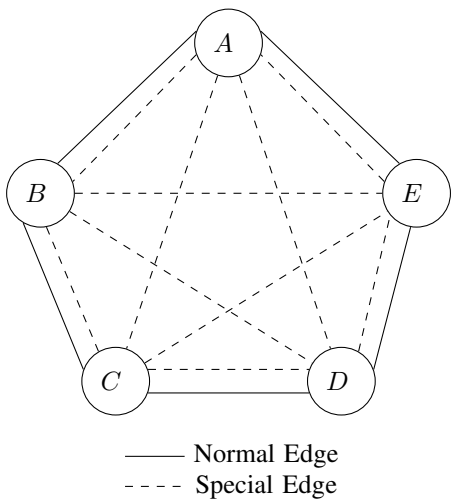
\begin{figure}
\centering
\resizebox{!}{!}{\begin{tikzpicture}

%\tikzset{
\tikzstyle{networknode} = [
%  networknode/.style = {
    draw,
    shape=circle,
%    radius=2cm,
    minimum width=0.9cm,
    text width=0.4cm
%  }
    ]
%}

%\DrawCircle{0}{0}{thick}{14pt}{black}{M}
%\DrawCircle{20}{0}{thick}{14pt}{black}{G}
%\DrawNormalLine{0}{10}{0}{12}{thick}
 %\DrawDashedLine{1}{10}{1}{12}{thick}
%\DrawDottedLine{2}{10}{2}{12}{thick}

%\newcommand\DoubleLine[2][{
%  \path(#1)--(#2)coordinate[at start](h1)coordinate[at end](h2);
%  \draw ($(h1)!90:(h2)$)--($(h2)!-90:(h1)$);
%  \draw[dashed] ($(h1)!-90:(h2)$)--($(h2)!90:(h1)$);
%}

\node[networknode] (A) at (0,0) {$A$};
\node[networknode] (B) at (-2.5,-2) {$B$};
\node[networknode] (C) at (-1.5,-4.5) {$C$};
\node[networknode] (D) at (1.5,-4.5) {$D$};
\node[networknode] (E) at (2.5,-2) {$E$};

\draw (A.160) -- (B.60);
\draw (B.240) -- (C.160);
\draw (C.340) -- (D.200);
\draw (D.20) -- (E.260);
\draw (E.120) -- (A.20);

\draw[dashed] (A.200) -- (B.20);
\draw[dashed] (B.280) -- (C.120);
\draw[dashed] (C.20) -- (D.160);
\draw[dashed] (D.60) -- (E.220);
\draw[dashed] (E.160) -- (A.340);

%\DoubleLine{A}{B}
%\DoubleLine{B}{C}
%\DoubleLine{C}{D}
%\DoubleLine{D}{E}
%\DoubleLine{E}{A}

\draw[dashed] (A) -- (C);
\draw[dashed] (A) -- (D);
\draw[dashed] (B) -- (D);
\draw[dashed] (B) -- (E);
\draw[dashed] (C) -- (E);

\node (blank1) at (-1.5, -5.5) {};
\node[right of = blank1, node distance = 2cm] (normaledge) {\text{Normal Edge}};
\draw (blank1) -- (normaledge);
\node[below of = blank1, node distance = 0.4cm] (blank2) {};
\node[right of = blank2, node distance = 2cm] (specialedge) {\text{Special Edge}};
\draw[dashed] (blank2) -- (specialedge);

\end{tikzpicture}}
\caption{Core Ring}
\label{CoreRing}
\end{figure}

The multi layer graph corresponding to a part of the Metro Network given in the previous section
is shown in Fig.~\ref{MultiLayerGraph}. To satisfy a service request between network elements \emph{s} and \emph{l} for
a SDH service layer rate VC12, the path that has to be computed should involve the network elements \emph{s}, \emph{r},
\emph{q}, \emph{e}, \emph{d}, \emph{l}. One of the path that has to be computed is \emph{S-s-s}, \emph{S-e-s}, \emph{S-m-t},
\emph{S-e-l}, \emph{R-e-l}, \emph{R-m-t}, \emph{R-e-s}, \emph{R-e-l}, \emph{Q-e-l}, \emph{E-e-l}, \emph{E-e-s},
\emph{E-s-s}, \emph{E-s-l}, \emph{E-o-d}, \emph{E-w-c}, \emph{E-w-l}, \emph{D-w-l}, \emph{D-w-c}, \emph{D-o-d},
\emph{D-s-l}, \emph{L-s-l}, \emph{L-s-s}. This represents a path such that the given SDH service is carried over a Circuit
Emulation Service (CES) between the network elements \emph{s} and \emph{e} and then as a SDH service between \emph{e} and \emph{l}.
The CES is adapted over a MPLS Tunnel between \emph{s} and \emph{r}. The SDH service between \emph{e} and \emph{l} is
adapted over OTN which is carried by a optical channel (OCH) over the WDM link between \emph{e} and \emph{d}. Finally,
the SDH link between \emph{d} and \emph{l} is used.

\begin{figure}
\centering
\resizebox{!}{!}{\begin{tikzpicture}

%\tikzset{
\tikzstyle{networknode} = [
%  networknode/.style = {
    draw,
    shape=circle,
%    radius=2cm,
    minimum width=0.9cm,
    text width=0.4cm
%  }
    ]
%}

%\DrawCircle{0}{0}{thick}{14pt}{black}{M}
%\DrawCircle{20}{0}{thick}{14pt}{black}{G}
%\DrawNormalLine{0}{10}{0}{12}{thick}
 %\DrawDashedLine{1}{10}{1}{12}{thick}
%\DrawDottedLine{2}{10}{2}{12}{thick}

%\newcommand\DoubleLine[2][{
%  \path(#1)--(#2)coordinate[at start](h1)coordinate[at end](h2);
%  \draw ($(h1)!90:(h2)$)--($(h2)!-90:(h1)$);
%  \draw[dashed] ($(h1)!-90:(h2)$)--($(h2)!90:(h1)$);
%}

\node[networknode] (F) at (0,0) {$F$};
\node[networknode] (G) at (-2,-2) {$G$};
\node[networknode] (H) at (0,-4) {$H$};
\node[networknode] (B) at (2,-2) {$B$};

\draw (F) -- (G);
\draw (G) -- (H);
\draw (H.345) -- (B.270);
\draw (B.90) -- (F.2);

\draw[dashed] (B.120) -- (F.330);
\draw[dashed] (B.240) -- (H.30);
\draw[dashed] (B) -- (G);

\node (blank1) at (-1, -5) {};
\node[right of = blank1, node distance = 2cm] (normaledge) {\text{Normal Edge}};
\draw (blank1) -- (normaledge);
\node[below of = blank1, node distance = 0.4cm] (blank2) {};
\node[right of = blank2, node distance = 2cm] (specialedge) {\text{Special Edge}};
\draw[dashed] (blank2) -- (specialedge);

%\DoubleLine{A}{B}
%\DoubleLine{B}{C}
%\DoubleLine{C}{D}
%\DoubleLine{D}{E}
%\DoubleLine{E}{A}

\end{tikzpicture}}
\caption{Ring with Aggregate Node}
\label{RingWithAggNode}
\end{figure}
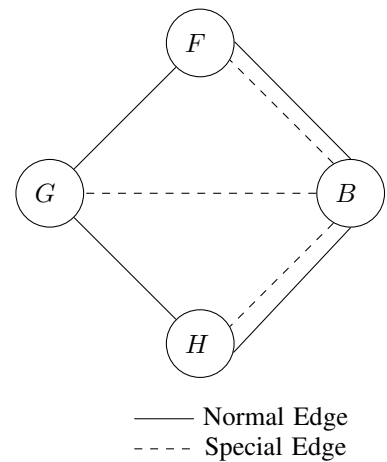

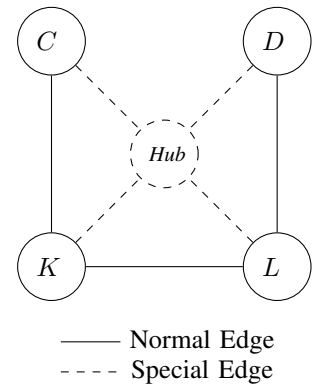
\begin{figure}
\centering
\resizebox{!}{!}{\begin{tikzpicture}

%\tikzset{
\tikzstyle{networknode} = [
%  networknode/.style = {
    draw,
    shape=circle,
%    radius=2cm,
    minimum width=0.9cm,
    text width=0.4cm
%  }
    ]
%}

%\DrawCircle{0}{0}{thick}{14pt}{black}{M}
%\DrawCircle{20}{0}{thick}{14pt}{black}{G}
%\DrawNormalLine{0}{10}{0}{12}{thick}
 %\DrawDashedLine{1}{10}{1}{12}{thick}
%\DrawDottedLine{2}{10}{2}{12}{thick}

%\newcommand\DoubleLine[2][{
%  \path(#1)--(#2)coordinate[at start](h1)coordinate[at end](h2);
%  \draw ($(h1)!90:(h2)$)--($(h2)!-90:(h1)$);
%  \draw[dashed] ($(h1)!-90:(h2)$)--($(h2)!90:(h1)$);
%}

\node[networknode] (C) at (0,0) {$C$};
\node[networknode] (D) at (3,0) {$D$};
\node[networknode] (K) at (0,-3) {$K$};
\node[networknode] (L) at (3,-3) {$L$};
\node[networknode, dashed] (Hub) at (1.5,-1.5) {\footnotesize \textit{Hub}};

\draw (C) -- (K);
\draw (K) -- (L);
\draw (L) -- (D);

\draw[dashed] (C) -- (Hub);
\draw[dashed] (D) -- (Hub);
\draw[dashed] (K) -- (Hub);
\draw[dashed] (L) -- (Hub);

\node (blank1) at (0, -4) {};
\node[right of = blank1, node distance = 2cm] (normaledge) {\text{Normal Edge}};
\draw (blank1) -- (normaledge);
\node[below of = blank1, node distance = 0.4cm] (blank2) {};
\node[right of = blank2, node distance = 2cm] (specialedge) {\text{Special Edge}};
\draw[dashed] (blank2) -- (specialedge);

%\DoubleLine{A}{B}
%\DoubleLine{B}{C}
%\DoubleLine{C}{D}
%\DoubleLine{D}{E}
%\DoubleLine{E}{A}

\end{tikzpicture}}
\caption{Dual Homing Ring}
\label{DualHomingRing}
\end{figure}

In the auxiliary graph structure, special edges are created to represent topologies in the network.
When there are parallel topologies like rings between a set of nodes in the network, the
working and protection paths should be computed such that the sub-paths belong to the same
set of topologies. This would help the operator in easy maintenance of circuits and
also reduce the problem troubleshooting time during failures. The topology name
could be set as an attribute of every physical or logical link in the network and during auto-discovery
of the network, the topologies could be derived and the auxiliary graph construction would
use that to build the topologies as special edges.

For the core ring (Fig.~\ref{CoreRing}) with no aggregate node involving the nodes \emph{a}, \emph{b}, \emph{c}, \emph{d} and \emph{e},
full mesh of special edges are added between every pair of nodes and the special edges are shown as dashed lines. This is to ensure that
the core ring is completely utilized during path computation. The normal links are shown as solid lines
indicating the actual connectivity.

For the aggregate ring (Fig.~\ref{RingWithAggNode}) involving the nodes \emph{b}, \emph{f}, \emph{g} and \emph{h} with
\emph{b} being the aggregate node, special edges are added from every node other than the
aggregate node to the aggregate node \emph{b} and the special edges are shown as dashed lines. This is to ensure that
the aggregate ring is completely utilized during path computation. The normal links are shown as solid lines
indicating the actual connectivity.

For the aggregate ring (Fig.~\ref{DualHomingRing}) involving the nodes \emph{c}, \emph{d}, \emph{k} and \emph{l} with \emph{c}
and \emph{d} being the aggregate nodes, special edges are added from every node other than the
aggregate nodes to a special node \emph{hub} and special edges are added from \emph{hub} node to the
aggregate nodes \emph{c} and \emph{d} and the special edges are shown as dashed lines. This is to ensure that
the dual homing ring is completely utilized during path computation. The normal links are shown as solid lines
indicating the actual connectivity.

%In all the cases where special edges are created for logical topologies, only physical and logical
%links are considered and the adaptation links are not considered for such a creation.

\subsection{Logical Link Creation}

Logical links are bearers in some particular layer or technology which rides over one or more physical
links and/or logical links in the server layer or technology. Logical links can multiplex and carry traffic from the
client layer or technology. Logical links are created in the following two scenarios:

\begin{enumerate}
\item{When there is some multiplexing or aggregation being done by means of server layer adaptation over
a series of nodes. For example, in SDH, VC4 is a multiplexing layer for many VC12, VC3 etc. signals and in WDM, OTU is
an aggregation layer to which multiple ODU signals can be mapped. In this case, the logical link is between the
terminating ports adjacent to the nodes at which those adaptation is performed. This corresponds to the
adaptation between layers within a technology.}
\item{When there is involvement of inter technology physical links in the end nodes which are used to
map traffic from one technology to another technology. In SDH or Ethernet over WDM, the SDH or Ethernet port
is connected by a physical link to a corresponding transponder port for that technology and rate for mapping into
a multiplexed WDM signal. In this case, the logical link is between the ports connected by inter technology links.
This corresponds to the adaptation across technologies.}
\end{enumerate}

\subsection{Weight Assignment}

In this section, the mechanism for assigning weights for different types of edges in the graph is
described. Weight assignment is crucial since the suitable paths for service requests can be found
by different weight functions for different types of edges. The initial weight setting mechanism for different
types of edges is described as follows:

\begin{itemize}
\item{\emph{Physical Links : }The initial weight for physical links has to be a function of link distance
and maximum capacity that could be supported on that link which is decided by the link type and link rate.
Link distance is used to avoid longer paths in the network. Maximum capacity is used to give preference
for links that support more capacity.
%Therefore, the weight for a physical link has to be directly
%proportional to the link distance and inversely proportional to the maximum capacity supported on that
%link which is specified in the below equation:

%\begin{equation}
%\label{eqn_phylinkwt_prop}
%w(pl) \propto d(pl) / b(pl)
%\end{equation}

%where $w(pl)$ is the weight of the physical link $pl$, $d(pl)$ is the link distance and $b(pl)$ is the
%maximum capacity supported on that link.

Setting the initial weight for physical links as a function of link distance and capacity of the link
gives a method for getting optimal paths for provisioning service requests that reduces the overall
utilization of the network. This is achieved by the weight assignment strategy proposed in the
below equation:

\begin{equation}
\label{eqn_phylinkwt}
w(L_p) = (\alpha * d(L_p) / d_{max}) + (\beta * (1 - b(L_p) / b_{max}))
\end{equation}

where $w(L_p)$ is the weight of the physical link $L_p$, $d(L_p)$ is the link distance, $d_{max}$ is the
maximum link distance across the network, $b(L_p)$ is the maximum capacity supported on the link $L_p$,
$b_{max}$ is the maximum capacity of any link across the network and $\alpha$ and $\beta$ are values
between 0 and 1 such that $\alpha + \beta = 1$. 

The above weight assignment strategy achieves normalization of weights for physical links across
the network based on their distance and capacity.}

\item{\emph{Logical Links : }The initial weight for logical links has to be a function of the sum
of the underlying physical or logical link weights so that it reflects both the link distance and the
maximum capacity of the physical or logical links involved. It also has to be less than the total sum of
the weights of the physical or logical links involved so that this logical link is preferred to the
series of the physical or logical links involved. This is specified in the below equation:

\begin{equation}
\label{eqn_loglinkwt}
w(L_l) = \gamma * \sum\limits_{l\in P} w(l)
\end{equation}

where $w(L_l)$ is the weight associated with the logical link $L_l$, $w(l)$ is the weight
associated with the physical or logical link $l$ involved, $P$ is set of links on the path through which
this logical link is established, $\gamma$ is some value between 0 and 1.}

\item{\emph{Special Edges for ring topology : }The initial weight for the special edges from each node
in that ring topology to the aggregate node of the ring topology has to be a function of the sum of
the weights of the physical or logical links involved in forming the ring topology.
%This is because the ring is self
%protected which means the working path will reach the aggregate node is one direction and the protection
%path will reach the aggregate node in the other direction thus involving all the links in that ring
%topology.
It also has to be less than the total sum of the weights of the physical and logical links
involved in the ring so that the special edge is preferred over the series of links for reaching the
aggregate node. This is specified in the below equation:

\begin{equation}
\label{eqn_ringtopowt}
w(T_r) = \eta * \sum\limits_{l\in P} w(l)
\end{equation}

where $w(T_r)$ is the weight associated with the special edge $T_r$ between each node in the ring to the
aggregate node, $w(l)$ is the weight associated with the physical or logical link $l$ involved
in forming the ring, $P$ is set of links on the ring topology, $\eta$ is some value between 0 and 1.}

\item{\emph{Special Edges for dual homing topology : }The initial weight for the special edges from each node
in that dual homing topology to the hub node of the dual homing topology has to be a function of the sum of
the physical or logical links involved in forming the dual homing topology.
%This is because the dual homing
%ring is self protected which means the working path will reach one aggregate node is one direction and the protection
%path will reach the other aggregate node in the other direction through the hub node thus involving all the
%links in that dual homing topology.
It also has to be less than the total sum of the weights of the physical and logical links
involved in the dual homing ring so that the special edge is preferred over the series of links for reaching the
aggregate nodes. This is specified in the below equation:

\begin{equation}
\label{eqn_dualhomingtopowt}
w(T_d) = \eta * \sum\limits_{l\in P} w(l)
\end{equation}

where $w(T_d)$ is the weight associated with the special edge $T_d$ between each node in the dual homing ring to the
hub node, $w(l)$ is the weight associated with the physical or logical link $l$ involved
in forming the dual homing ring, $P$ is set of links on the dual homing ring, $\eta$ is some value between 0 and 1.
For the special edges between the hub node and the two aggregate nodes in the dual homing ring, the initial weight
can be zero since the involved link weights are accounted in the special edges from each node to the hub node.}

\end{itemize}

\subsection*{Dynamic Weight Assignment}

Using only the initial static weight for provisioning could result in blocking of future service
requests due to continuous usage of capacity in that part of the network.
Dynamic weight assignment is required to distribute the capacity requirements
in the network instead of over-utilizing the same part of the network.
Following three different dynamic weight assignment mechanisms are proposed and evaluated in this
work.

\textbf{1. Piece-wise Linear Function based on Utilization (PLF)}

Based on the current utilization of links in the network, weights have to be dynamically varied
so that links with lesser utilization are preferred over links with high utilization.
The following is one of the dynamic weight assignment strategy which is a piece-wise linear function
based on the approach proposed in \cite{PEFT:Xu}:

\begin{enumerate}
\item{If $\rho_{cur}$ $\leq (1/3)$ then $W_d = W_i + W_{prop}$.}
\item{Else if $(1/3) <$ $\rho_{cur}$ $\leq (2/3)$ then $W_d = 2 * W_i + W_{prop}$}
\item{Else if $(2/3) <$ $\rho_{cur}$ $\leq (9/10)$ then $W_d = 5 * W_i + W_{prop}$}
\item{Else if $\rho_{cur}$ $> (9/10)$ then $W_d = 10 * W_i + W_{prop}$}
\end{enumerate}

In the above scheme, $\rho_{cur}$ refers to the current utilization of a link,
$W_i$ and $W_d$ refer to the initial weight and dynamic weight
respectively for a link. $W_{prop}$ refers to the proportional increase in weight
based on the current utilization of a link.

\textbf{2. Logarithmic Function based on Distance and Utilization (LF)}

\begin{equation}
  D(L_p)=\begin{cases}
    w(L_p),                                             & \text{if $f(L_p) = b(L_p)$}\\
    \alpha * d(L_p) - \\
    (\beta * \log(f(L_p) / b(L_p))),  & \text{otherwise}
  \end{cases}
\end{equation}

%\begin{equation}
%\label{eqn_lf}
%\left.
%D(L_p) = 
%\{
%\right.
%%\begin{aligned}
%w(L_p) \quad if \quad f(L_p) = b(L_p)\\
%\alpha * d(L_p) - (\beta * log(f(L_p) / b(L_p))) \quad if \quad f(L_p) \neq b(L_p)
%\end{aligned}
%\end{equation}

This dynamic weight function is based on two parameters of a physical link namely distance and utilization.
It is based on the Hop Count, Total and Available Wavelength (HTAW) weight assignment proposed
in \cite{bhide2001routing}. It computes the weight as a negative logarithm of utilization
since the weights are additive in nature.
%This dynamic weight function is as follows:

In the above scheme, $D(L_p)$ is the dynamic weight of the physical link $L_p$,
$w(L_p)$ is the initial weight of the physical link $L_p$, $d(L_p)$ is the link distance,
$f(L_p)$ is the free capacity in the link $L_p$, $b(L_p)$ is the maximum capacity
supported on the link $L_p$ and $\alpha$ and $\beta$ are values
between 0 and 1 such that $\alpha + \beta = 1$.

\textbf{3. Weighted Geometric Mean of Distance and Utilization (WGM)}

\begin{equation}
  D(L_p)=\begin{cases}
    1,                                                         & \text{if $\alpha = 0, \rho_{cur} = 0$}\\
    w(L_p),                                                    & \text{if $\rho_{cur} = 0$}\\    
    \exp(\alpha * \log(d(L_p)) + \\
    (\beta * \log(\rho_{cur}))),  & \text{otherwise}
  \end{cases}
\end{equation}

This dynamic weight function is based on two parameters of a physical link namely distance and utilization
where the weighted geometric mean of link distance and current utilization is computed.
Weighted geometric mean is used since it can be used to combine the two parameters of a link namely
distance and utilization without normalizing them.
%This dynamic weight function is as follows:

In the above scheme, $D(L_p)$ is the dynamic weight of the physical link $L_p$,
$w(L_p)$ is the initial weight of the physical link $L_p$, $d(L_p)$ is the link distance,
$\rho_{cur}$ refers to the current utilization of that link and $\alpha$ and $\beta$ are values
between 0 and 1 such that $\alpha + \beta = 1$.

\section{Path Computation Algorithm}

%In this section, the algorithm for path computation over the auxillary graph structure proposed
%in the earlier section is described. 

Path computation in multi-layer networks is shown to be NP-Complete in \cite{kuipers2009path} and \cite{PCMLAlgoComp:Lamali}. Hence,
heuristic path computation algorithm is proposed
using the proposed auxiliary graph structure.
The algorithm takes as input the source and destination
nodes which are obtained from the corresponding network elements and the layerRate for which
the path is required. For example, to compute the path between two network elements for a SDH
VC12 layerRate, the nodes corresponding to SDH Technology and Service layer will be selected.
The algorithm also takes as input the pathType which could be unprotected or link-disjoint path
pair (fully protected), the number of paths (N) required and the layerRate and capacity (in case
of Ethernet). Any request for a path for some service would be associated with a bandwidth or capacity
which has to be available in all the edges of the computed path.
For PDH or SDH like TDM interface port, at the end nodes of the service request,
the bandwidth or capacity is generally specified as layerRate (ex. VC12 in SDH means 2 Mbps).
This is taken as the input to path computation since even though the requested
capacity may be available in a link but the capacity as a single rate container
may not be available. For example, capacity equivalent to VC3 in SDH (45 Mbps) may be available
but not as a single VC3 continer.

Even though competitive analysis is generally used to analyze online
algorithms, it has not been used in this work since topology
constraints (working and protection path follow same set of
topologies) which is a practical requirement in service provider
networks to satisfy customer service requests, initial and dynamic
weight assignment for links that is based on their distance and
utilization, are considered. Hence, in such a complex practical setting
which is specific to transport networks, computing an optimal offline
algorithm for comparing with the online version for competitive
analysis is not practical.

\begin{algorithm}
\caption{findPath(source, dest, pathType, N, layerRate, capacity)}
\label{pathcomputealgo}
\begin{algorithmic}[1]
\STATE Delete the edges whose available capacity is less than the layerRate and the capacity requested
\IF{pathType = UNPROTECTED\_PATH}
\STATE \emph{pathList} = findUnprotectedPath(source, dest, N)
\ELSE
\IF{pathType = LPP}
\STATE \emph{pathList} = findLPP(source, dest, N)
\ENDIF
\ENDIF
\IF{$N == 1$}
%\FOR{\emph{path} in \emph{pathList}}
\STATE \emph{path} = \emph{pathList}[0]
\STATE checkAndCreateLogicalLinks(\emph{path})
%\STATE Remove the edges for which equivalent logical links are created in the above step and replace them with the corresponding logical links in the correct position
\STATE Provision the requested capacity for the given layerRate in the logical links created and the other links
\STATE Subtract the provisioned capacity from the available capacity in the logical links created and the other links
\ENDIF
\STATE Restore the edges deleted in step 1
\RETURN \emph{pathList}
\end{algorithmic}
\end{algorithm}

\begin{algorithm}
\floatname{algorithm}{Procedure}
\caption{findUnprotectedPath(source, dest, N)}
\label{unprotectedpathcomputealgo}
\begin{algorithmic}[1]
\STATE Remove the edges including special edges whose self-protected flag is set to true
\IF{$N == 1$}
\STATE \emph{path} = findDijkstraShortestPath(source, dest)
\IF{\emph{path} != $NULL$}
\STATE add \emph{path} to \emph{pathList}
\ENDIF
\ELSE
\STATE \emph{pathList} = findKShortestLooplessPaths(source, dest, N)
\ENDIF
\STATE Restore the edges deleted in step 1
\RETURN \emph{pathList}
\end{algorithmic}
\end{algorithm}

%\begin{figure}
%  \centering
%  \resizebox{0.6\textwidth}{0.6\textwidth}{\input{UnprotectedPathAlgo.tikz}}
%  \label{UnprotectedPathAlgo}
%\end{figure}

The algorithm for path computation (Algorithm \ref{pathcomputealgo}) takes as input the
source and destination nodes, the path type which could be unprotected path or dedicated
protection path (LPP), the number of paths to be found - N, the layer rate for the requested
service (ex. VC12, Ethernet Bandwidth etc.) and the capacity in Mbps if the layer rate is Ethernet
Bandwidth. It removes the edges which do not have enough capacity to meet the requested capacity.
It calls the algorithm for finding the unprotected path or dedicated protection path which are outlined in
Procedure \ref{unprotectedpathcomputealgo} and Procedure \ref{lpppathcomputealgo} respectively.
These algorithms return the list of paths computed depending on the number of
paths to be found, N. If the number of paths requested is 1, then the best path is given
to the Procedure \ref{logicallinkcreation} to check and create logical links. The requested
capacity for the given layer rate is provisioned in the logical links created and the other
physical and logical links in the path found and the available capacity is subtracted from the
provisioned capacity in those links. If the number of paths requested is greater than 1, then
the user has to select the path to be used and the logical link creation for that path, provisioning the
requested capacity in that path and the available capacity adjustment in that path has to be invoked.
Finally, the deleted edges are restored back to the graph and the list of paths found is returned by
the algorithm.

The algorithm for unprotected path computation (Procedure \ref{unprotectedpathcomputealgo})
takes as input the source and destination nodes, and the number of paths to be found - N.
It removes the edges including special edges which are self-protected since the requirement is to
find unprotected path so that the protection part of the self-protected edge is not wasted.
If the number of paths requested is 1, the algorithm calls the Dijkstra shortest path algorithm
to find the shortest path between the given source and destination nodes. Else, it calls the Yen's K Shortest
path algorithm \cite{KShortAlgo:Yen} to find the best N paths between the given source and destination nodes. Finally,
the deleted edges are restored back to the graph and the list of paths found is returned by
the algorithm.

\emph{Computational Complexity:} The computational complexity of
Procedure \ref{unprotectedpathcomputealgo} in the average case is
$\Theta(((E+V)\log V))$ if $N$ is equal to 1 and
$\Theta((NV(E+V\log V)))$ if $N$ is not equal to 1
where $N$ is the number of paths requested, $V$ and $E$ are the number of nodes and
edges in the auxiliary graph. If $N$ is equal to 1, then the complexity of the algorithm is the
complexity of Step 3 which is the complexity of the Dijkstra'a shortest path algorithm.
If $N$ is not equal to 1, then the complexity of the algorithm is the complexity of
Step 7 which is the complexity of the Yen's K Shortest path algorithm.

\begin{algorithm}
\floatname{algorithm}{Procedure}
\caption{findLPP(source, dest, N)}
\label{lpppathcomputealgo}
\begin{algorithmic}[1]
\STATE Remove the involved links in all the topologies since special edges only have to be considered
\STATE \emph{pathList} = findKShortestLooplessPaths(source, dest, N)
\FOR{$i$ in $1 \ldots{}$ \emph{pathList.size()}}
\STATE \emph{workingPath} = pathList.get(i)
\STATE Remove the edges in \emph{workingPath} other than adaptation edges for protection path computation
\STATE Remove the edges involved in the same SRLGs for edges from \emph{workingPath}
\STATE $protectionPathFound \leftarrow \TRUE$
\STATE Break the \emph{workingPath} into segments that are either self-protected or not without considering adaptation edges
\FOR{each segment \emph{segment} in \emph{workingPath}}
\IF{\emph{segment} is not self-protected}
\STATE \emph{path} = findUnprotectedPath(source of \emph{segment}, dest of \emph{segment}, 1)
\IF{\emph{path} != $NULL$}
\STATE Add \emph{path} to \emph{protectionPath}
\ELSE
\STATE $protectionPathFound \leftarrow \FALSE$
\STATE break
\ENDIF
\ENDIF
\ENDFOR
\STATE Restore the edges deleted in step 4 and 5
\IF{$protectionPathFound \neq \TRUE$}
\STATE continue;
\ENDIF
\STATE Add \emph{workingPath} and \emph{protectionPath} to \emph{pathList}
\ENDFOR
\STATE Sort the \emph{pathList} based on total cost of \emph{workingPath} and \emph{protectionPath}
\RETURN sorted \emph{pathList}
\end{algorithmic}
\end{algorithm}

%\begin{figure}
%  \centering
%  \resizebox{!}{!}{\input{LPPAlgo.tikz}}
%  \resizebox{0.8\textwidth}{0.8\textwidth}{\input{LPPAlgo.tikz}}
%  \label{LPPAlgo}
%\end{figure}

%\subsection{Algorithm for LPP Computation}

%\begin{algorithm}
%\caption{exploreNode(node, currentRate, stack)}
%\label{explorenode}
%\begin{algorithmic}[1]
%\IF{\emph{node} supports cross-connection at \emph{currentRate}}
%\RETURN true
%\STATE Let \emph{allowedCCOrAdaptationRateSet} represent cross-connection or adaptation allowed rates in \emph{node}
%\STATE \emph{clientRate} = \emph{currentRate}
%\WHILE{true}
%\STATE \emph{serverRate} represent the immediate server or adaptation allowed rate for \emph{clientRate}
%\IF{\emph{serverRate} is $NULL$}
%\STATE break
%\ENDIF
%\IF{\emph{serverRate} is present in \emph{allowedCCOrAdaptationRateSet}}
%\RETURN $(true,serverRate)$
%\ELSE
%\STATE add \emph{clientRate} to \emph{stack}
%\STATE \emph{clientRate} = \emph{serverRate}
%\ENDIF
%\ENDWHILE
%\WHILE{\emph{stack} is not empty}
%\STATE \emph{poppedRate} = pop(\emph{stack})
%\IF{\emph{poppedRate} is present in \emph{allowedCCOrAdaptationRateSet}}
%\RETURN $(true,poppedRate)$
%\ENDIF
%\ENDWHILE
%\RETURN $(false,NULL)$
%\end{algorithmic}
%\end{algorithm}

%\begin{algorithm}
%\caption{exploreEdge(edge, currentRate)}
%\label{exploreedge}
%\begin{algorithmic}[1]
%\IF{\emph{currentRate} is compatible with the type and rate of \emph{edge}}
%\RETURN true
%\ELSE
%\RETURN false
%\ENDIF
%\end{algorithmic}
%\end{algorithm}

%\subsection{Algorithm for Edge Exploration}

\begin{algorithm}
\floatname{algorithm}{Procedure}
\caption{exploreEdge(edge, currentNode, adjacentNode, stack)}
\label{exploreedge}
\begin{algorithmic}[1]
\IF{\emph{edge} is an adaptation edge}
\STATE Let $(Tech,Layer)$ correspond to the adaptation type in the top of the \emph{stack}
\STATE Let $(Tech1,Layer1)$ correspond to the adaptation type in the \emph{currentNode}
\STATE Let $(Tech2,Layer2)$ correspond to the adaptation type in the \emph{adjacentNode}
\IF{$(Tech,Layer)$ = $NULL$}
\STATE push $(Tech1,Layer1)$ to \emph{stack}
\ELSE
\IF{$(Tech1,Layer1)$ and $(Tech2,Layer2)$ correspond to server to client adaptation}
\IF{$(Tech,Layer)$ = $(Tech2,Layer2)$}
\STATE pop the \emph{stack}
\ELSE
\RETURN false
\ENDIF
\ELSE
\IF{$(Tech1,Layer1)$ and $(Tech2,Layer2)$ correspond to client to server adaptation}
\STATE push $(Tech1,Layer1)$ to \emph{stack}
\ENDIF
\ENDIF
\ENDIF
\ENDIF
\RETURN true
\end{algorithmic}
\end{algorithm}

The algorithm for protected path computation (Procedure \ref{lpppathcomputealgo})
takes as input the source and destination nodes, and the number of paths to be found N.
The first step is to remove all the involved edges in all the topologies since special edges only have to
be considered during LPP computation for topologies.
It invokes the Yen's K Shortest path algorithm \cite{KShortAlgo:Yen} to find the N working paths between the given
source and destination. The algorithm executes the number of paths found times iteratively. In the
$i^{th}$ iteration, protection path is found for segments in the working path that are not
self-protected. It removes the edges in the working path that are not adaptation edges and the edges
involved in the same SRLGs for the edges in the current working path. It breaks the current working
path into segments that are self-protected or not without considering adaptation edges. For each
segment that is not self-protected, the Procedure \ref{unprotectedpathcomputealgo} is invoked by
calling the segment start and end nodes as arguments and the number of paths to be found as 1 to
find the link-disjoint protection path for that segment. If protection path is found for all
the unprotected segments, then complete protection path is formed by assembling the protection
segments found and the working and protection paths are added to the path list.  Else, the
algorithm continues to the next iteration. At the end of all the iterations, the path list is
sorted based on the combined cost of the working and protection paths. The sorted path list
is finally returned by the algorithm.

\emph{Computational Complexity:} The computational complexity of
Procedure \ref{lpppathcomputealgo} in the average case is
$\Theta(NV((E+V\log V) + ((E+V)\log V)))$
where $N$ is the number of paths requested, $V$ and $E$ are the number of nodes and
edges in the auxiliary graph. The complexity of the Yen's K Shortest path algorithm
used in Step 1 is $\Theta((NV(E+V\log V)))$. The outer for loop in the algorithm is
executed N times. In each iteration of the outer for loop, the complexity of Step 7 is
$\Theta(V)$ and so the inner for loop is executed $V$ times. In each iteration of
the inner for loop, the complexity of Step 10 is $\Theta(((E+V)\log V))$ which is the
complexity of the Procedure \ref{unprotectedpathcomputealgo} with N set as 1. Hence
the complexity of the outer for loop is $\Theta(NV((E+V)\log V))$ and therefore the
complexity of the algorithm which is the sum of the complexities of Step 1 and the
outer for loop is $\Theta((NV(E+V)\log V))$.

\begin{algorithm}
\floatname{algorithm}{Procedure}
\caption{checkAndCreateLogicalLinks(path)}
\label{logicallinkcreation}
\begin{algorithmic}[1]
\STATE Initialize an empty \emph{stack}
\FOR{\emph{edge} in \emph{path}}
\IF{\emph{edge} is an adaptation edge}
\STATE Let $(Tech,Layer)$ correspond to the adaptation type in the top of the \emph{stack}
\STATE Let $(Tech1,Layer1)$ correspond to the adaptation type in the current node of \emph{edge}
\STATE Let $(Tech2,Layer2)$ correspond to the adaptation type in the next node of \emph{edge}
\IF{$(Tech,Layer)$ = $NULL$}
\STATE push $(Tech1,Layer1)$ to \emph{stack}
\ELSE
\IF{$(Tech1,Layer1)$ and $(Tech2,Layer2)$ correspond to server to client adaptation}
\IF{$(Tech,Layer)$ = $(Tech2,Layer2)$}
\STATE pop the \emph{stack}
\IF{$(Tech,Layer)$ do not correspond to a service layer}
\STATE Create a logical link between the nodes popped from the stack and the current node of \emph{edge}
\STATE Provision the capacity for the logical link in the involved physical and logical links
\STATE Subtract the provisioned capacity from the available capacity in the involved physical and logical links
\ENDIF
\ENDIF
\ENDIF
\IF{$(Tech1,Layer1)$ and $(Tech2,Layer2)$ correspond to client to server adaptation}
\STATE push $(Tech1,Layer1)$ to \emph{stack}
\ENDIF
\ENDIF
\ENDIF
\ENDFOR 
\end{algorithmic}
\end{algorithm}

\begin{figure*}
\centering
\subfloat[Weighted Number of Requests Accepted]{\includegraphics[width=0.45\linewidth]{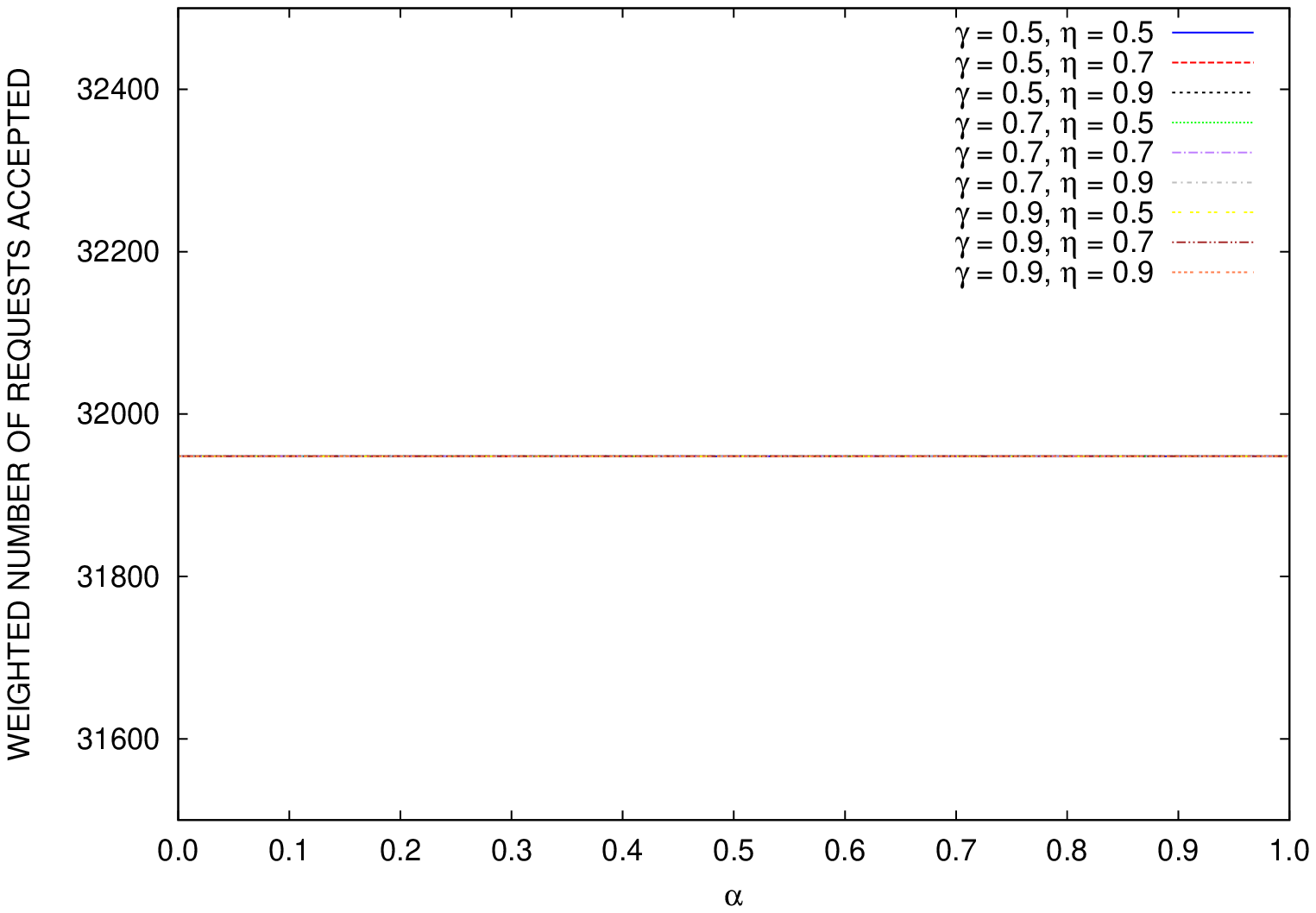}}
\subfloat[Total Bandwidth Consumed]{\includegraphics[width=0.45\linewidth]{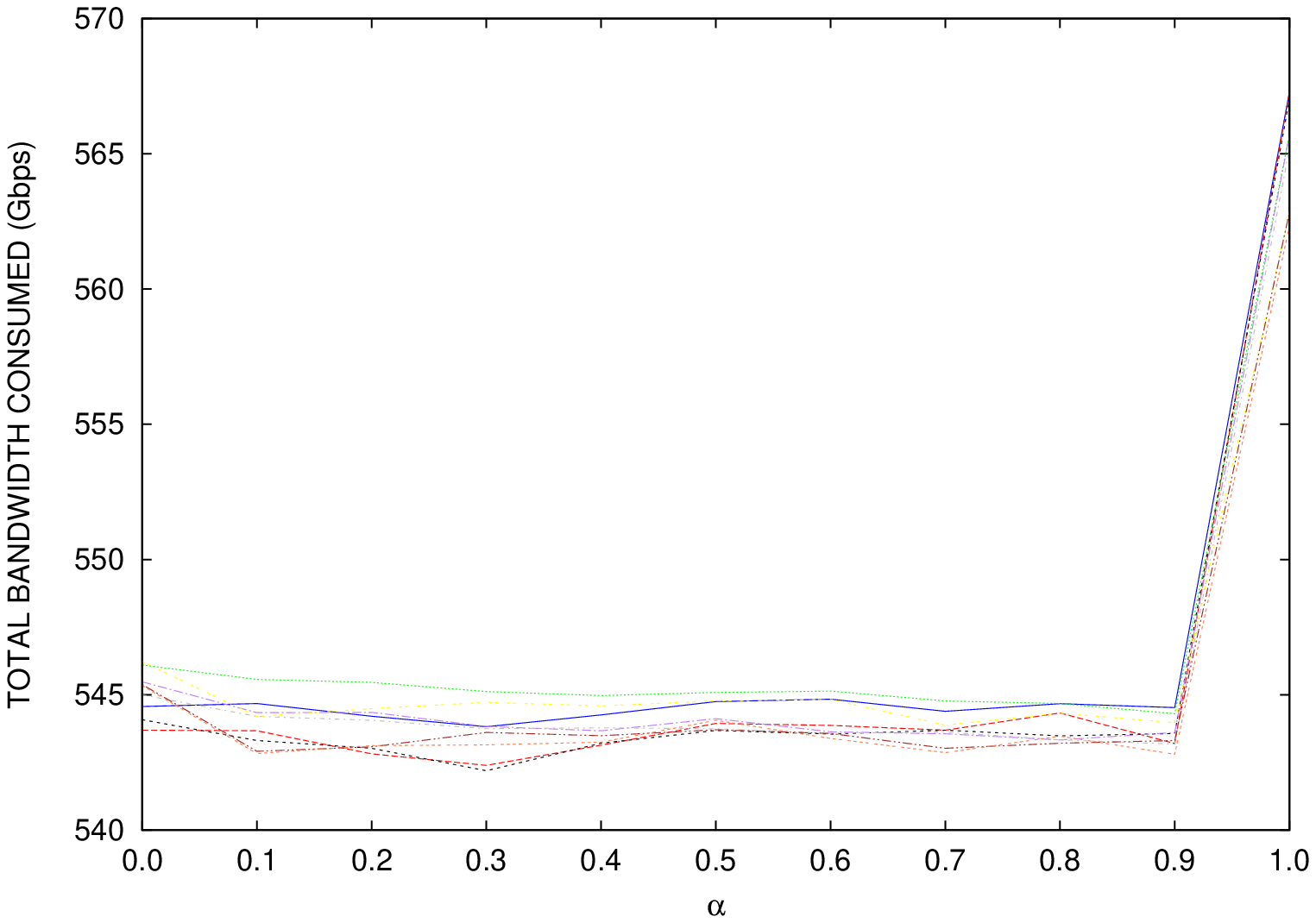}}

\subfloat[Total Number of Logical Links Created]{\includegraphics[width=0.45\linewidth]{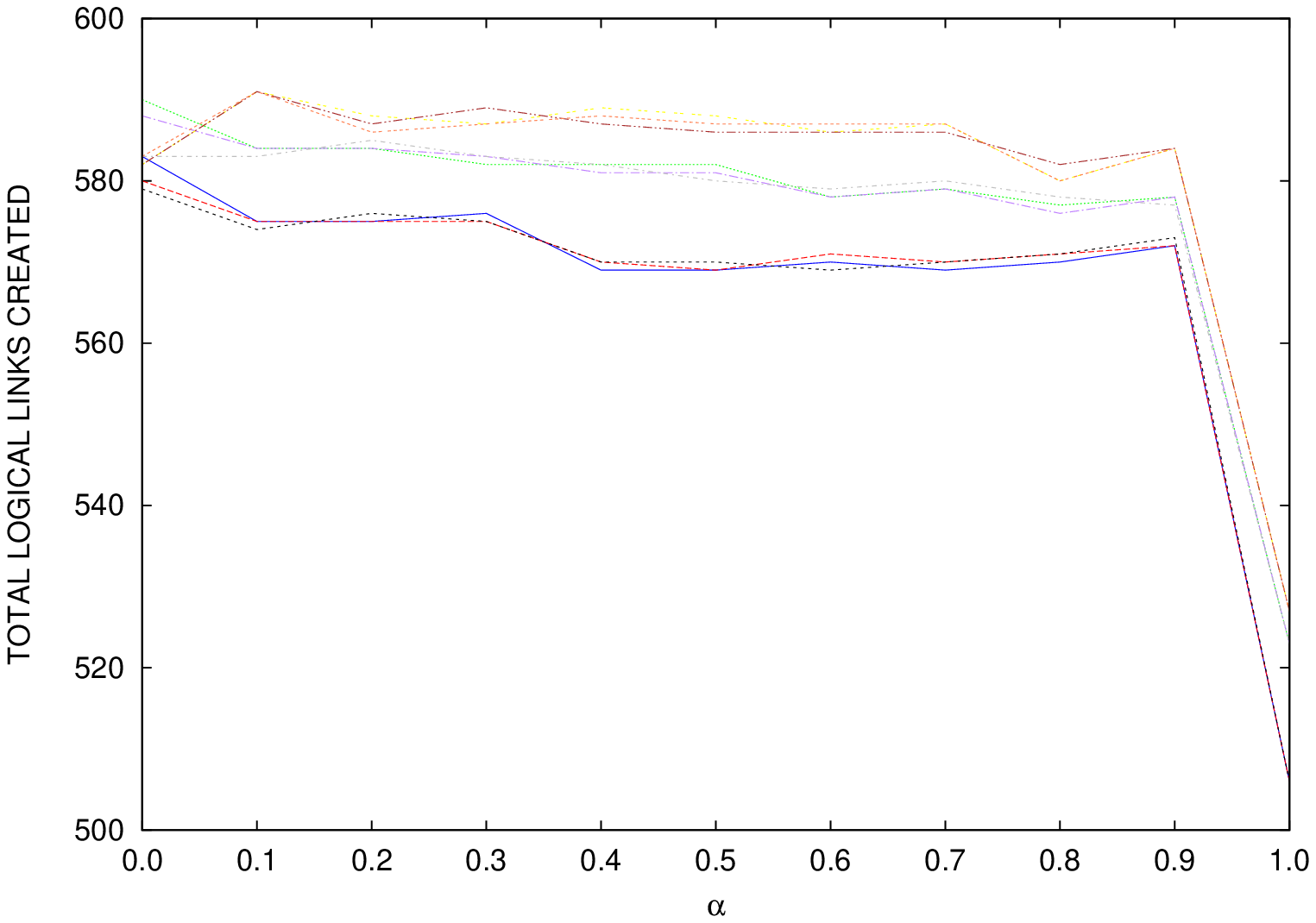}}
\subfloat[Weighted Link Utilization]{\includegraphics[width=0.45\linewidth]{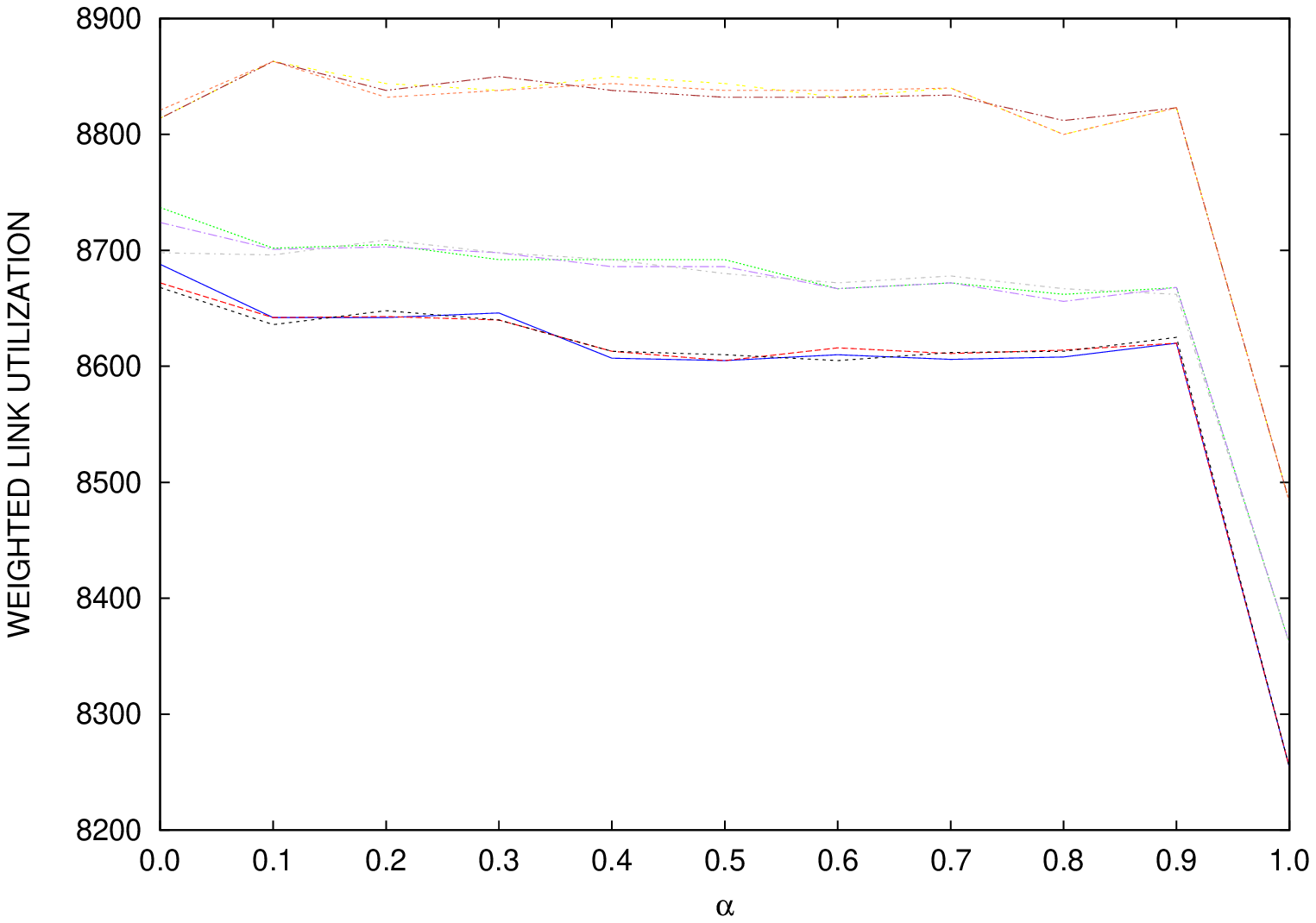}}
\caption{Performance Evaluation for PLF}
\label{PerfEvalPLF}
\end{figure*}

\begin{figure*}
\centering
\subfloat[Weighted Number of Requests Accepted]{\includegraphics[width=0.45\linewidth]{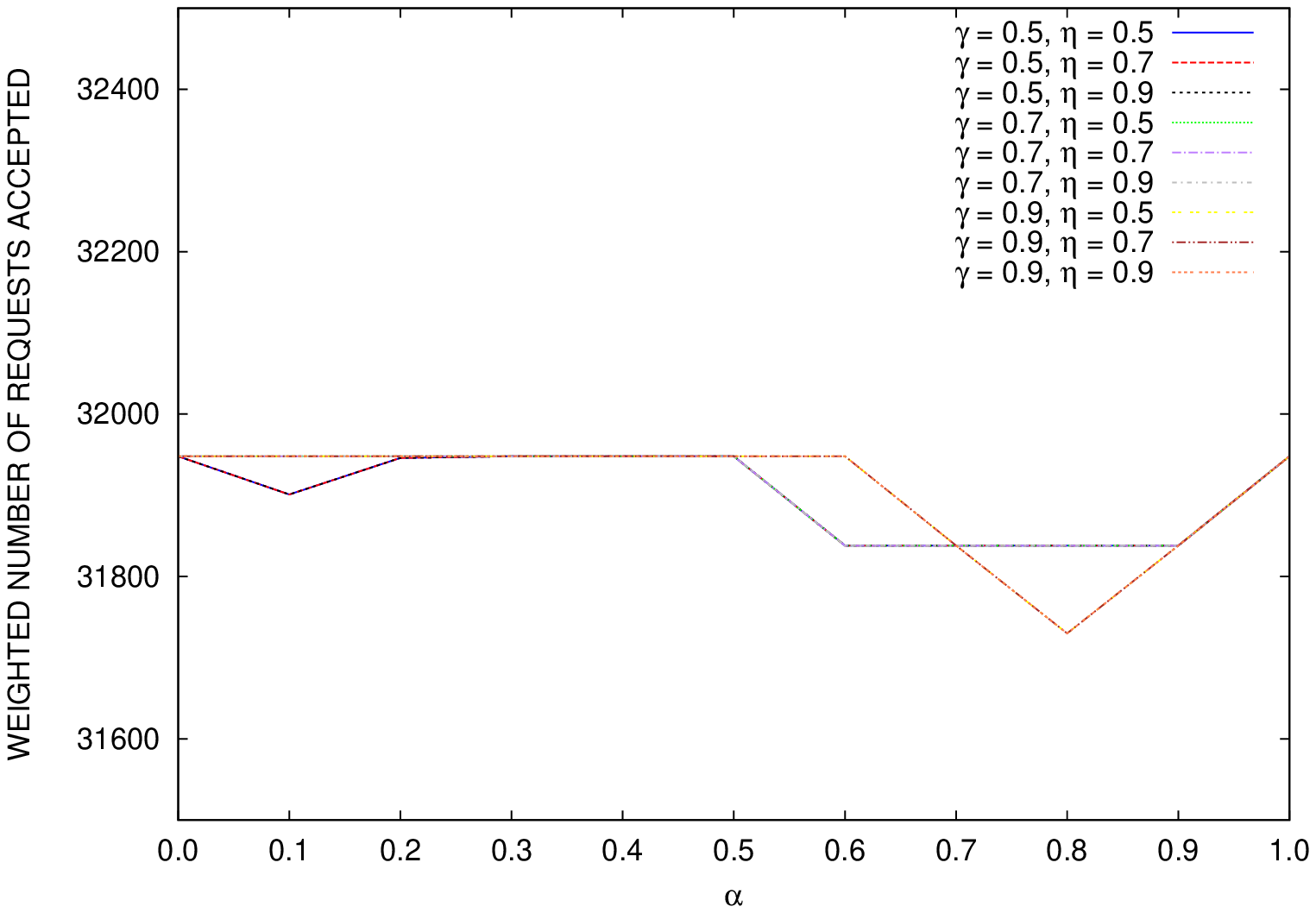}}
\subfloat[Total Bandwidth Consumed]{\includegraphics[width=0.45\linewidth]{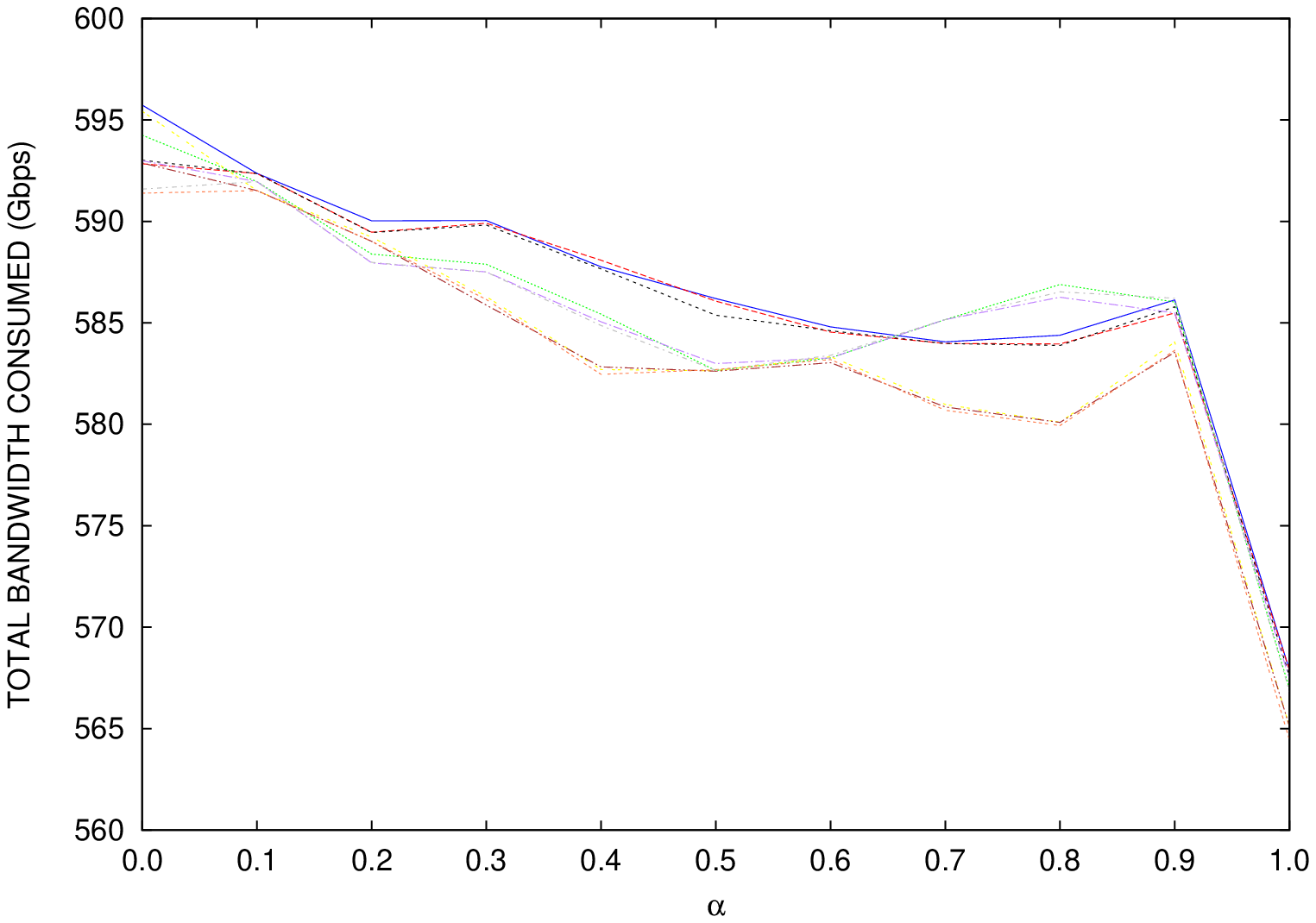}}

\subfloat[Total Number of Logical Links Created]{\includegraphics[width=0.45\linewidth]{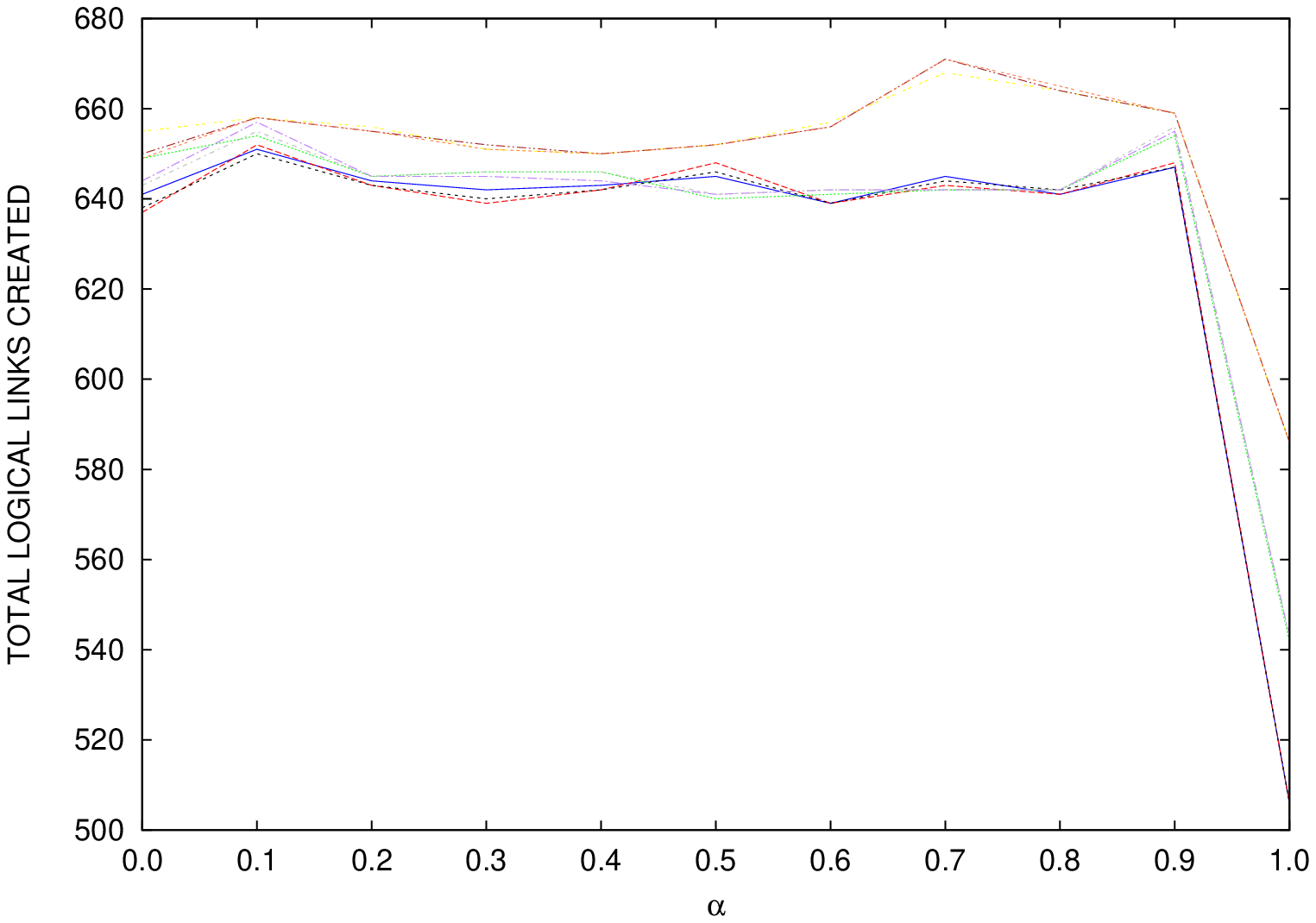}}
\subfloat[Weighted Link Utilization]{\includegraphics[width=0.45\linewidth]{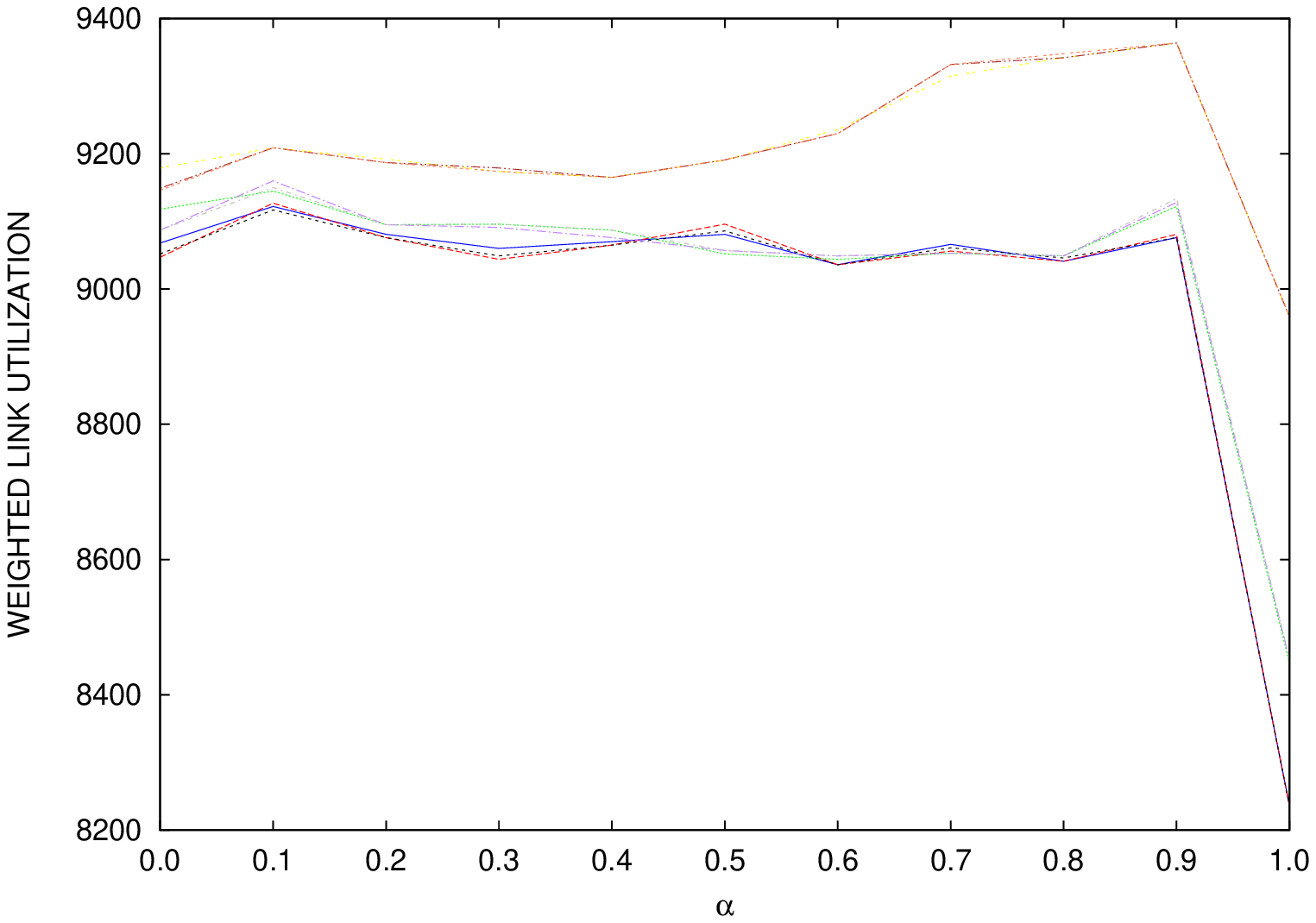}}
\caption{Performance Evaluation for LF}
\label{PerfEvalLF}
\end{figure*}

\begin{figure*}
\centering
\subfloat[Weighted Number of Requests Accepted]{\includegraphics[width=0.45\linewidth]{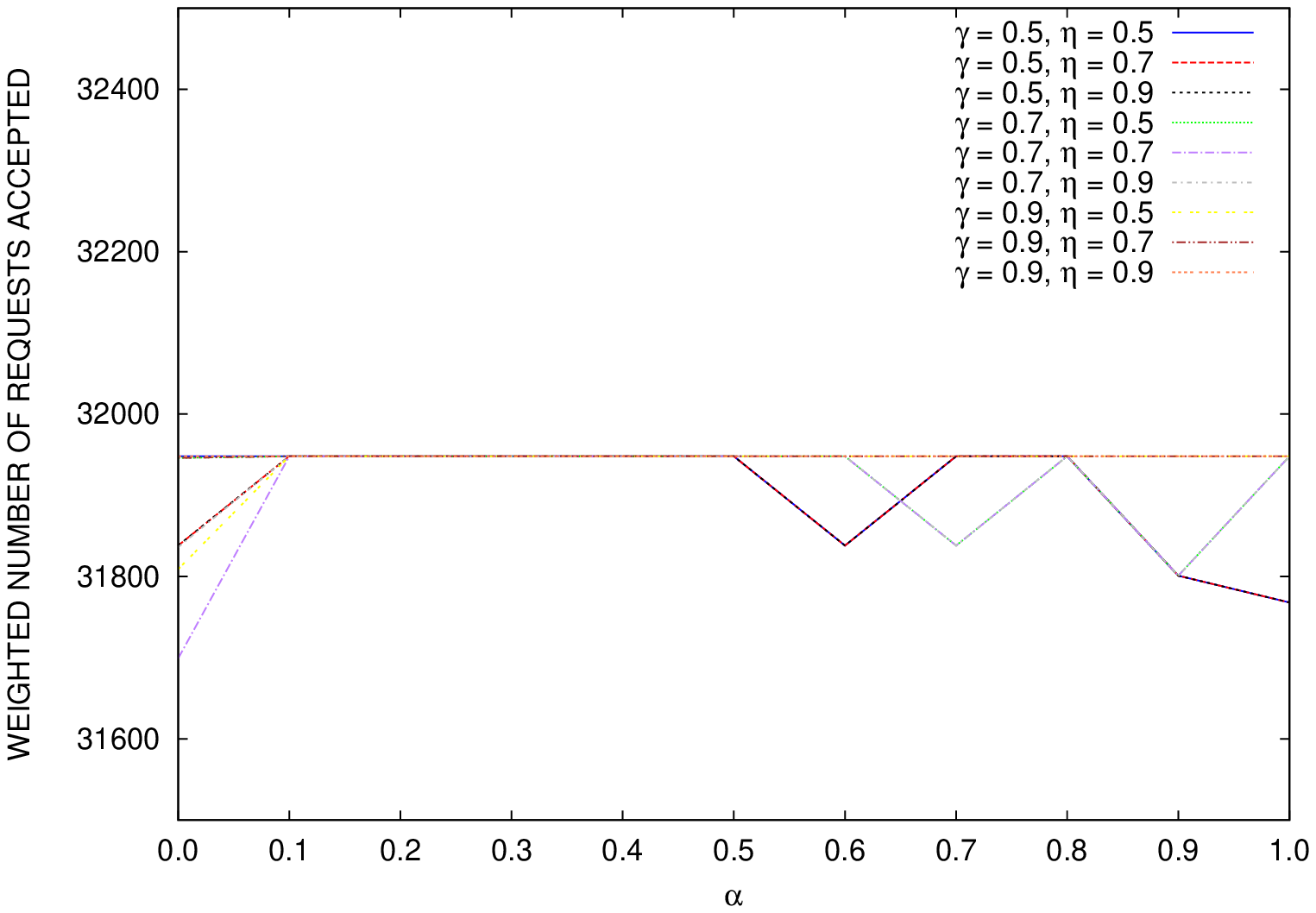}}
\subfloat[Total Bandwidth Consumed]{\includegraphics[width=0.45\linewidth]{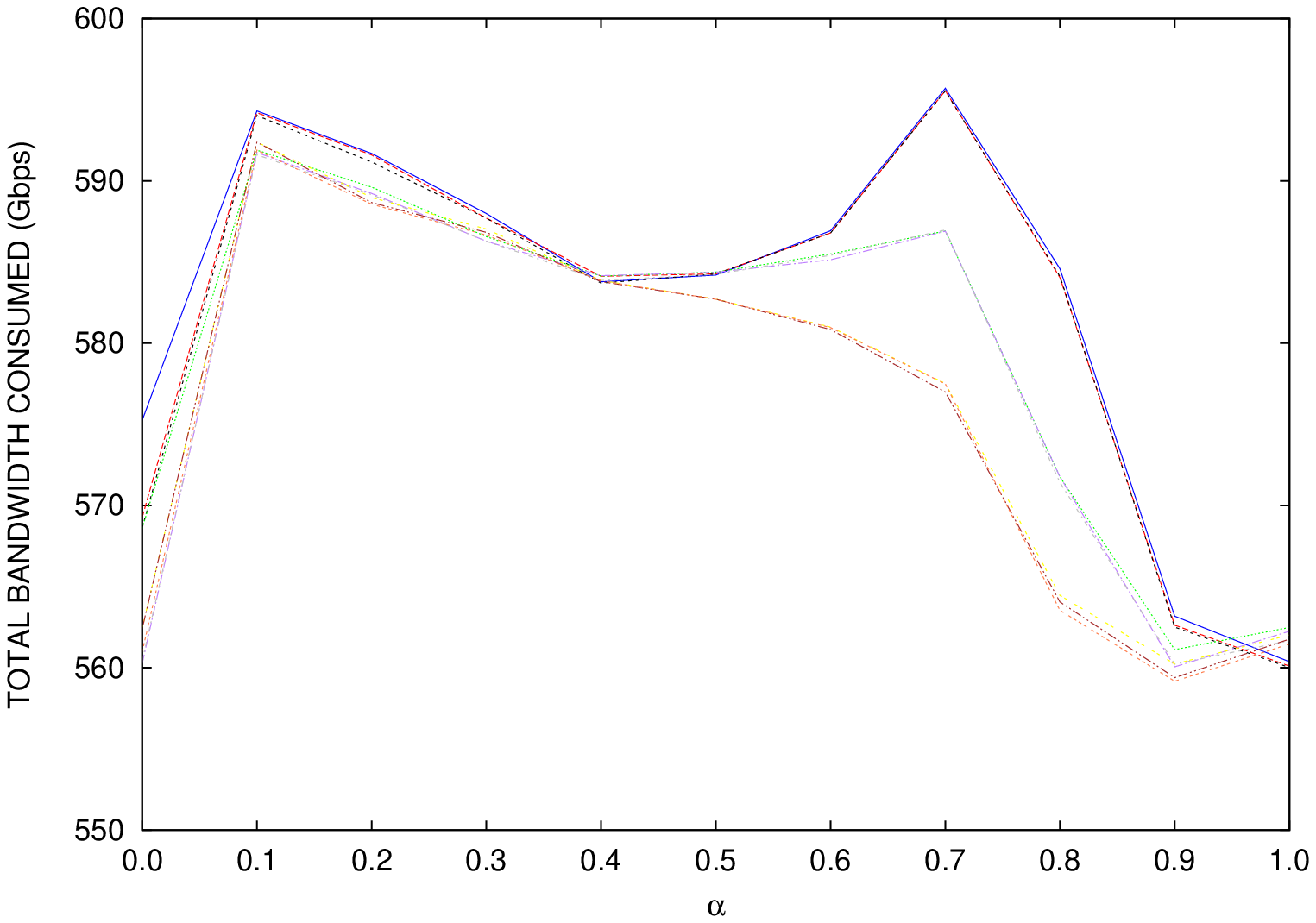}}

\subfloat[Total Number of Logical Links Created]{\includegraphics[width=0.45\linewidth]{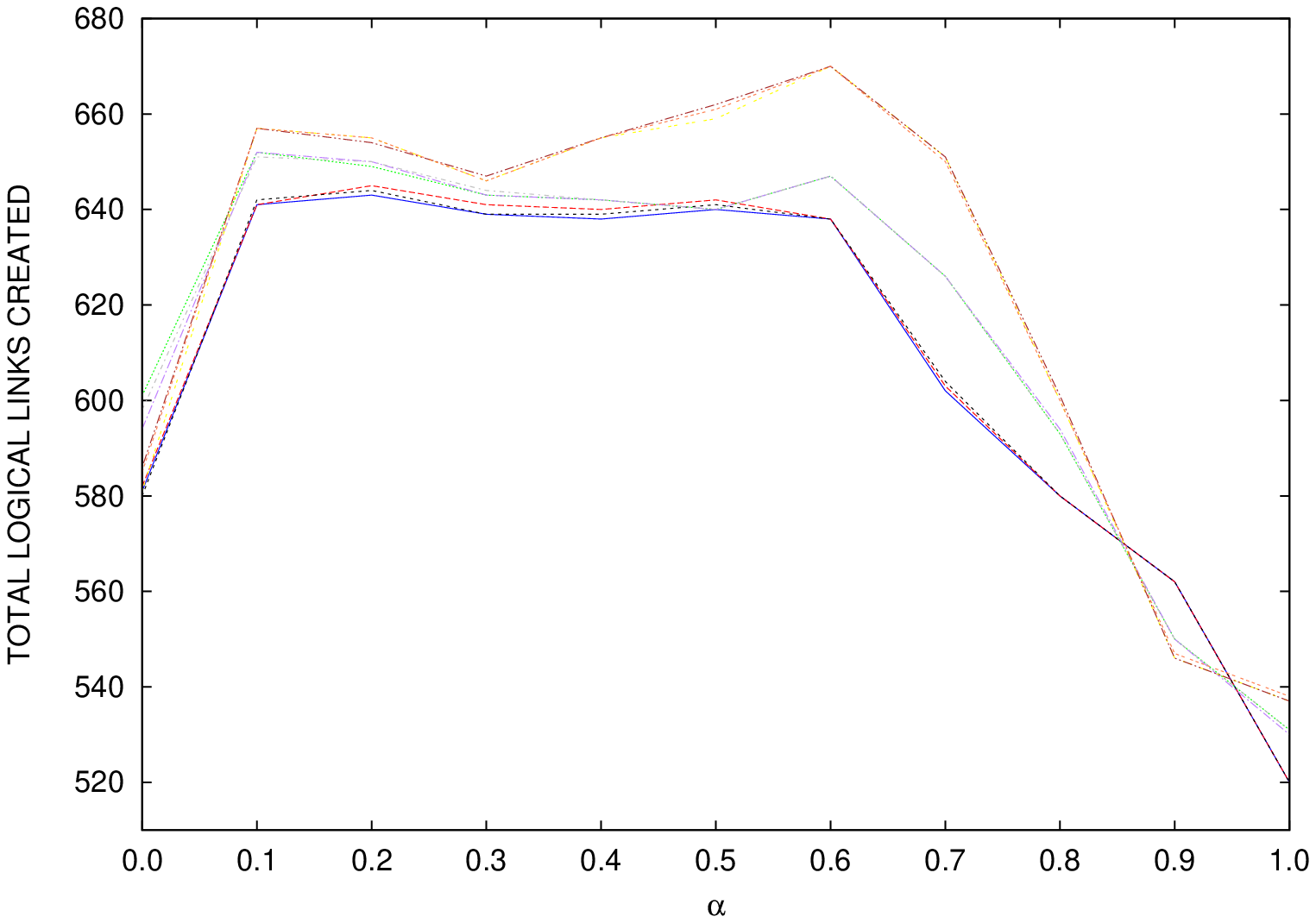}}
\subfloat[Weighted Link Utilization]{\includegraphics[width=0.45\linewidth]{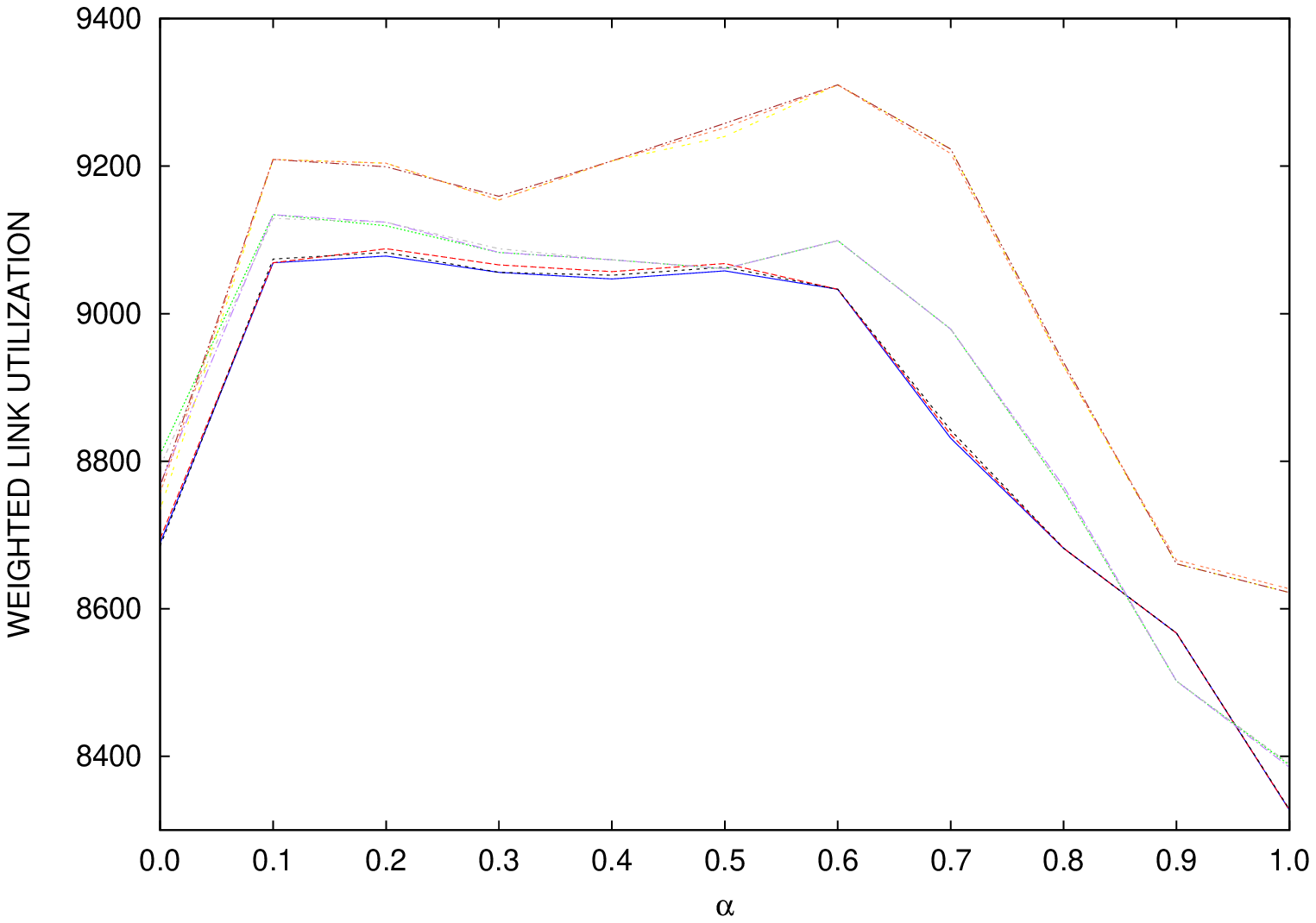}}
%\subfloat[Weighted Number of Requests Accepted]{\label{WGMWtdReq}\resizebox{!}{!}\includegraphics[width=0.45\textwidth]{WtdReqAvgLPPNew-WtApp3.eps}}
%\subfloat[Total Bandwidth Consumed]{\label{WGMWtdBWProv}\resizebox{!}{!}\includegraphics[width=0.45\textwidth]{WtdBWAvgLPPNew-WtApp3.eps}}
%
%\subfloat[Total Number of Logical Links Created]{\label{WGMTotalLL}\resizebox{!}{!}\includegraphics[width=0.45\textwidth]{TotalLLAvgLPPNew-WtApp3.eps}}
%\subfloat[Weighted Link Utilization]{\label{WGMWtdLinkUtil}\resizebox{!}{!}\includegraphics[width=0.45\textwidth]{WtdLinkUtilAvgLPPNew-WtApp3.eps}}
\caption{Performance Evaluation for WGM}
\label{PerfEvalWGM}
\end{figure*}

During the path computation using Dijkstra's or Yen's algorithms in the above given procedures, 
adaptation edges have to be properly explored. For example, to satisfy a Ethernet service over SDH
link, the Ethernet service has to be adapted into SDH service which can be carried over the SDH link.
The SDH traffic may in turn get adapted into OTN. While leaving the OTN part of the network, the OTN
node may have capability to terminate both SDH and Ethernet traffic. Since in our example, SDH traffic
was carried over OTN, OTN to SDH adaptation edge has to be used instead of OTN to Ethernet adaptation
edge while leaving the last OTN node.

To ensure selection of the correct adaptation edges during path
computation, the Procedure \ref{exploreedge} is used. A stack of technology and layer
combination is maintained at each node during path computation. The stack at the source node is
initialized to null. While exploring each edge during path computation, if the edge is an adaptation
edge, then the technology and layer component of the node at the top of the stack, the current node
and the adjacent node are set to three different variables. If the stack is null, then the technology
and layer component of the current node is pushed onto the stack. Else, if the technology and layer
component of the current and adjacent nodes correspond to server to client adaptation (ex. OTN to SDH)
and the technology and layer component at the top of the stack equals that of the adjacent node, then
stack is popped else the edge exploration is returned as false to avoid incompatible adaptation edge
to be explored. If the technology and layer component of the current and adjacent nodes correspond to
client to server adaptation (ex. SDH to OTN), then the technology and layer component of the current
node is pushed onto the stack. Finally, the algorithm returns true to explore the current
adaptation edge.

%\subsection{Algorithm for Logical Link Creation}

%\begin{figure}
%  \centering
%  \includegraphics[width=0.45\textwidth]{WtdLinkUtilAvgLPPNew-WtApp3.eps}
%  \caption{chumma}
%  \label{chumma}
%\end{figure}

%\begin{figure*}[t]
%\begin{multicols}{2}
%    \includegraphics[width=\linewidth]{WtdReqAvgLPPNew-WtApp3.eps}\par\caption{caption}
%    \includegraphics[width=\linewidth]{WtdBWAvgLPPNew-WtApp3.eps}\par\caption{caption}
%\end{multicols}
%\begin{multicols}{2}
%    \includegraphics[width=\linewidth]{TotalLLAvgLPPNew-WtApp3.eps}\par\caption{caption}
%    \includegraphics[width=\linewidth]{WtdLinkUtilAvgLPPNew-WtApp3.eps}\par\caption{caption}
%\end{multicols}
%\caption{Performance Evaluation for WGM}
%\label{PerfEvalWGM}
%\end{figure*}

During path computation for a service request from one technology and layer, inter-technology links could
be used which results in creation of logical links. The procedure to check
and create logical links from the path found is outlined in Procedure \ref{logicallinkcreation}.
Similar to the Procedure \ref{exploreedge} that explore edges during path computation, this algorithm
also maintains a stack to track of client to server and server to client adaptations. Whenever the client
to server adaptation (ex. SDH to OTN) is encountered, it pushes the client technology and layer
combination onto the stack. When the corresponding server to client adaptation is encountered, it pops the
stack and the logical link is created between the client technology and layer at the corresponding nodes.
While creating the logical link, the capacity corresponding to the logical link is provisioned in the server
layer links and the available capacity in the server layer links is adjusted.

\section{Performance Results}

The performance of the described algorithms using the weight assignment schemes
proposed are evaluated and the results obtained are provided in this section.
A typical service provider network scenario is considered for performance evaluation.
The network consists of national long distance (NLD) portion with 15 locations
for nodes and 23 links connecting them. The links support WDM with 80 channels
capacity where each channel can carry OTN signal of OTU-2 rate. The OTU-2 signal
carry SDH or Ethernet traffic.

In each location part of the NLD network, a metro network is formed. The metro network
in a location contains a metro core mesh network with 4 nodes and links connecting
every pair of nodes that support WDM with 40 channels. Two aggregate normal rings with 2 nodes
and 2 links each of STM-64 and 10 GE rate are formed
and are connected to the metro core node in that location using 1+1 protected links
of STM-64 and 10 GE rate. In each node of the aggregate ring, normal access rings
with 2 nodes and 2 links each of STM-16 and GE rate are formed.
Also, two aggregate dual homing rings with 4 nodes and links each of STM-64 and
10 GE rate are formed and are connected to the metro core node
in that location using 1+1 protected links of STM-64 and 10 GE rate.
In each node of the dual homing aggregate ring, normal access rings
with 4 nodes and links each of STM-16 and GE rate are formed.

In each location, random aggregate rings with already used 4 random nodes and links
each of STM-64 and 10 GE rate are formed
and are connected to the metro core node in that location using 1+1 protected links
of STM-64 and 10 GE rate. The links between metro core node and the NLD node in every location
are formed using 1+1 protection of STM-64 and 10 GE rate.

This network formation resulted in 2955 network elements which in turn resulted in
10455 nodes (due to creation of nodes for each technology and layer combination
within every network element), 480 hub nodes for dual homing topologies and
10380 adaptation edges, 5393 physical links and 3540 special edges for a total of
19313 edges in the complete auxiliary graph.

To evaluate the performance for finding LPP paths, 500 service requests
between randomly selected source and destination nodes in the above network
are generated. The service requests are generated at SDH layer with rate
VC12, VC3 and VC4 and Ethernet layer with random capacity between 1 to 200 Mbps.
%The process is repeated for 6 iterations where in each iteration the value
The process is executed such that the value
for $\alpha$ is varied between 0.0 to 1.0 in increments of 0.1 such that
the value for $\beta$ is set as $1 - \alpha$ and the value for $\gamma$ and $\eta$ are
varied between 0.5 to 0.9 in increments of 0.2.

\begin{table}
\centering
\subfloat[Weight Approach]{
\begin{tabular}{|l|r|}
\hline
Approach & Value\\
\hline
PLF & 0.3316\\
LF & 0.3329\\
WGM & 0.3325\\
\hline
\end{tabular}}%\hfill
\subfloat[ALPHA]{
\begin{tabular}{|l|r|}
\hline
ALPHA & Value\\
\hline
0.0 & 0.3324\\
0.1 & 0.3324\\
0.2 & 0.3324\\
0.3 & 0.3323\\
0.4 & 0.3323\\
0.5 & 0.3323\\
0.6 & 0.3327\\
0.7 & 0.3328\\
0.8 & 0.3326\\
0.9 & 0.3326\\
1.0 & 0.3309\\
\hline
\end{tabular}}%\hfill

\subfloat[GAMMA]{
\begin{tabular}{|l|r|}
\hline
GAMMA & Value\\
\hline
0.5 & 0.3322\\
0.7 & 0.3323\\
0.9 & 0.3325\\
\hline
\end{tabular}}%\hfill
\subfloat[ETA]{
\begin{tabular}{|l|r|}
\hline
ETA & Value\\
\hline
0.5 & 0.3323\\
0.7 & 0.3324\\
0.9 & 0.3323\\
\hline
\end{tabular}}
\caption{Combined Performance of 4 Parameters}
\label{combinedperf}
\end{table}

The tests were performed on a machine with 4-core Intel i5 processor (3 GHz) and 64 GB RAM. 
For each combination of $\alpha$, $\gamma$ and $\eta$, the weighted number of requests accepted
, the total capacity utilized in the complete network, the total number of logical
links that are created and the weighted link utilization are found.
Weighted number of requests accepted is used since the bandwidth of each
service request accepted is not the same and hence the requests accepted
are weighted by their bandwidth requirement.
For calculating the weighted link utilization, the links in the network are categorized into
four types based on their percentage utilization; links utilized less than or equal to 25\%, links utilized
greater than 25\% and less than or equal to 50\%, links utilized
greater than 50\% and less than or equal to 75\% and links utilized
greater than 75\% and less than or equal to 100\%. The number of links falling under each category
is multiplied by the factors 1, 2, 3 and 4 (to reflect their relative utilization)
for the above four types respectively resulting in the weighted link utilization.
%For each combination of
%$\alpha$, $\gamma$ and $\eta$, the total amount of capacity utilized in the complete
%network is found.

The time taken for unprotected path computation in the auxiliary graph
for the mentioned network was 400 milliseconds on average for each service
request with a maximum of 900 milliseconds. For the LPP service requests, the time
taken was 15 seconds on average with a maximum of 40 seconds.

The results obtained from one of the iterations is shown in
the figures Fig.~\ref{PerfEvalPLF}, Fig.~\ref{PerfEvalLF} and Fig.~\ref{PerfEvalWGM}
(legends for the individual figures (b), (c) and (d) are same as that of figure (a)
and are omitted for better visual clarity)
for the three dynamic weight assignment mechanisms proposed.
For each dynamic weight function, the four parameters mentioned earlier are evaluated.
From the results in the graphs, the performance of PLF weight function
is better in terms of more weighted number of requests accepted, lesser total capacity
utilized, lesser total number of logical links created and lesser weighted link
utilization compared to the WGM weight function which is in turn better compared
to the LF weight function. For the weighted number of requests accepted,
the difference in obtained values in not that
significant whereas for the other three parameters the difference is comparatively more.

The combined performance of the four parameters is then evaluated by computing the
average of the other three parameters relative to the weighted number of requests accepted.
This combined performance is computed as follows:

\begin{equation}
\label{combinedperfeqn}
\begin{split}
val & = ((B_{w\alpha\gamma\eta} / R_{w\alpha\gamma\eta}) * (1 + \log(B_{w\alpha\gamma\eta}^{max} / B_{w\alpha\gamma\eta}))) \\
&\quad + ((L_{w\alpha\gamma\eta} / R_{w\alpha\gamma\eta}) * (1 + \log(L_{w\alpha\gamma\eta}^{max} / L_{w\alpha\gamma\eta}))) \\
&\quad + ((U_{w\alpha\gamma\eta} / R_{w\alpha\gamma\eta}) * (1 + \log(U_{w\alpha\gamma\eta}^{max} / U_{w\alpha\gamma\eta})))
\end{split}
\end{equation}

where $w$ represents the weight approach (PLF, LF, WGM), $R_{w\alpha\gamma\eta}$ represents
the weighted number of requests accepted, $B_{w\alpha\gamma\eta}$ represents the total capacity
utilized, $L_{w\alpha\gamma\eta}$ represents the total number of logical links created,
$U_{w\alpha\gamma\eta}$ represents the weighted link utilization for the combination of $w$,
$\alpha$, $\gamma$, $\eta$ respectively and $B_{w\alpha\gamma\eta}^{max}$ represents the maximum of total
capacity utilized, $L_{w\alpha\gamma\eta}^{max}$ represents the maximum of total number of
logical links created and $U_{w\alpha\gamma\eta}^{max}$ represents the maximum of weighted link
utilization across all values of $w$, $\alpha$, $\gamma$, $\eta$ respectively. The value $val$
obtained is averaged across all values of $w$, $\alpha$, $\gamma$, $\eta$ individually and is
shown in the Table \ref{combinedperf}. The lower value indicates better performance and it can
be observed that PLF performs better than WGM which is in turn better than LF.
It can be observed that for values of $\alpha$ between 0.3 and 0.5,
the performance is better (0.5 being the best) for the four parameters considered.
This clearly shows that a combination of link distance and link capacity
for weight assignment gives better results for LPP computation.
Also, among the three values 0.5, 0.7 and 0.9 for $\gamma$ and $\eta$, for the four
parameters considered, there is not much significant difference among them (0.5 gives
the lowest value and better performance) and 
it can be taken that the logical link weight
and the special edge weight has to be lesser than the sum of the
involved link weights.
 
%Also, among the three values 0.5, 0.7 and 0.9 for $\gamma$ and $\eta$, for the four
%parameters considered, the value of 0.7 gave better performance across the three dynamic
%weight functions. This shows that the logical link weight
%and the special edge weight has to be only slightly lesser than the sum of the
%involved link weights.

%The combined performance of the four parameters across the three weight functions
%for different combinations of $\alpha$, $\gamma$ and $\eta$ is observed to show similar results
%and is not shown here due to lack of space.
%is shown in the Table \ref{relativefactor}.
%The combined performance for each combination is computed by taking the obtained value
%relative to the maximum value for each parameter and their sum is taken. 
%Lower the value indicates better performance and it can be observed that
%the PLF has better performance compared to LF which in turn has better performance than WGM.

\section{Conclusions}

In this paper, the problem of path computation for service provisioning in
transport networks is addressed. An auxiliary graph structure that models
the different characteristics of the transport network using different
technologies is proposed. Adaptation of traffic from one technology and layer
to another is modeled as adaptation edges in the auxiliary graph.
Since the transport networks are organized as topologies like ring, dual
homing etc. special edges that model these characteristics are created in
the auxiliary graph. Weight assignment for links such that both link distance
and link capacity are considered, is proposed. Dynamic weight assignment for
links that consider the current utilization of the links in the graph is
also proposed to prefer lesser utilized links over highly utilized
links.

Heuristic path computation algorithms for finding unprotected and link-disjoint
path pair over the auxiliary graph structure using the proposed weight assignment
schemes are described. Support for creating logical links upon encountering
inter-technology boundary is provided so that future service requests could use them.
The algorithms and the weight assignment schemes are evaluated and from the results
obtained it can be observed that the combination of link distance and link capacity
for initial weight assignment performs better when compared to consideration of
only one of those factors. Also, it is found that the dynamic weight function
PLF performs better when compared to the function LF which in turn is better than the function WGM.
%Piece-wise Linear Function based on Utilization (PLF)
%performs better when compared to the weight function 
%Logarithmic Function based on Distance and Utilization (LF) which in turn is better
%than the third weight function Weighted Geometric Mean of Distance and Utilization (WGM).
Studying the behaviour of the proposed algorithm over time
and optimization of the network with services provisioned is planned as a future work.

%\bibliographystyle{./IEEEtran}
%\bibliography{./IEEEabrv,./IEEEexample,./paper}
\bibliographystyle{abbrv}
\bibliography{pcatn-paper}

%\begin{IEEEbiography}[{\includegraphics[width=1in,height=1.25in,clip,keepaspectratio]{madan-color.eps}}]{Madanagopal Ramachandran}
\begin{IEEEbiographynophoto}{Madanagopal Ramachandran}
received the B.E. degree in computer science and engineering from Bharathiar
University, Coimbatore, India, in 2002 and the M.S. degree in computer
science and engineering from the Indian Institute of Technology, Madras,
India, in 2008. He is currently a Senior Technical Architect with NMSWorks Software
Pvt. Ltd., Chennai, India.
%He is currently leading the design and
%development of a product named Optical Transport Network Management
%System (OTNMS). He has worked on the project Advanced Network Systems
%in Industrial Consultancy and Sponsored Research, Indian Institute of
%Technology, Madras, India. His research interests include service
%provisioning in transport networks, networking, algorithms and
%distributed systems.
%\end{IEEEbiography}
\end{IEEEbiographynophoto}

%\begin{IEEEbiography}[{\includegraphics[width=1in,height=1.25in,clip,keepaspectratio]{KS.S-2012.eps}}]{Krishna M. Sivalingam}
\begin{IEEEbiographynophoto}{Krishna M. Sivalingam}
is a Professor and Head in the Department of CSE, IIT
Madras, Chennai, INDIA. He received his Ph.D. and M.S. in Computer
Science from SUNY Buffalo; and his B.E.  degree from Anna University's
College of Engineering Guindy, India.  His research interests include
wireless and optical networks. He is an IEEE Fellow, INAE Fellow and
ACM Distinguished Scientist.
%Until recently, he served as
%Editor-in-Chief of Springer Photonic Network Communications Journal
%and EAI Transactions on Future Internet.

%From 1994-2007, he was a faculty member in University of Maryland,
%Baltimore County; Washington State University, Pullman and University
%of North Carolina Greensboro, all in the USA.
%\end{IEEEbiography}
\end{IEEEbiographynophoto}

\end{document}